\documentclass[american,aps,pre,reprint,showpacs,superscriptaddress,altaffilletter]{revtex4-1}
\usepackage{lmodern}
\usepackage{lmodern}
\usepackage[T1]{fontenc}
\usepackage[utf8]{inputenc}
\usepackage{color}
\usepackage{babel}
\usepackage{array}
\usepackage{bm}
\usepackage{amsmath}
\usepackage{amssymb}
\usepackage{graphicx}
\usepackage[unicode=true,
 bookmarks=true,bookmarksnumbered=false,bookmarksopen=false,
 breaklinks=false,pdfborder={0 0 0},pdfborderstyle={},backref=false,colorlinks=true]
 {hyperref}
\hypersetup{
 pdfborderstyle={},pdfborderstyle={},pdfborderstyle={},pdfborderstyle={},linkcolor=blue,citecolor=blue}

\makeatletter

\providecommand{\tabularnewline}{\\}

 \@ifundefined{textcolor}{}
 {%
   \definecolor{BLACK}{gray}{0}
   \definecolor{WHITE}{gray}{1}
   \definecolor{RED}{rgb}{1,0,0}
   \definecolor{GREEN}{rgb}{0,1,0}
   \definecolor{BLUE}{rgb}{0,0,1}
   \definecolor{CYAN}{cmyk}{1,0,0,0}
   \definecolor{MAGENTA}{cmyk}{0,1,0,0}
   \definecolor{YELLOW}{cmyk}{0,0,1,0}
 }

\usepackage{babel}
\usepackage{babel}
\usepackage{babel}
\usepackage{babel}
\usepackage{comment}

\newcommand{\sn}{\,\mathrm{sn}}
\newcommand{\cn}{\,\mathrm{cn}}
\newcommand{\dn}{\,\mathrm{dn}}
\newcommand{\E}{E}
\newcommand{\K}{K}

\makeatother

\begin{document}

\title{Fluid-supported elastic sheet under compression: Multifold solutions}

\author{Leonardo Gordillo}
\email{leonardo.gordillo@usach.cl}

\affiliation{Departamento de Física, Universidad de Santiago de Chile, \\
 Av. Ecuador 3493, Estación Central, Santiago, Chile}

\author{Edgar Knobloch}
\email{knobloch@berkeley.edu}

\affiliation{Department of Physics, University of California at Berkeley, Berkeley,
California 94720, USA}
\begin{abstract}
The properties of a hinged floating elastic sheet of finite length
under compression are considered. Numerical continuation is used to
compute spatially localized buckled states with many spatially localized
folds. Both symmetric and antisymmetric states are computed and the
corresponding bifurcation diagrams determined. Weakly nonlinear analysis
is used to analyze the transition from periodic wrinkles to single
fold and multifold states and to compute their energy. States with
the same number of folds have energies that barely differ from each
other and the energy gap decreases exponentially as localization increases.
The stability properties of the different competing states are established
and provide a basis for the study of fold interactions. 
\end{abstract}


\date{\today}

\maketitle

\section{Introduction}

Recent experiments \citep{Pocivavsek:2008kz} have shown that an elastic
sheet on the surface of a liquid undergoes a series of transitions
when compressed in the longitudinal direction: for small compression
the sheet deforms into a periodic array of wrinkles; with increasing
compression the deformation of the sheet becomes more and more nonlinear
and the deformation spontaneously localizes. This process continues
until the sheet makes contact with itself.

These experiments have stimulated several efforts at modeling this
sequence of transitions \citep{2011PhRvL.107p4302D,Audoly:2011cq,Rivetti:2014dh,Oshri:2015bi}.
The equilibrium configurations are described by a fourth-order equation
for the deformation angle as a function of arclength. This equation,
as well as others describing post-buckling phenomena of elastic media
under compression \citep{Hunt:D0NJhxAR} and torsion \citep{1996RSPSA.452..117T,1996RSPSA.452.2467C},
is of Swift-Hohenberg type and its solutions bear a number of similarities
to existing analysis of steady states of the Swift-Hohenberg equation
with a quadratic nonlinearity \citep{1997JFM...342..199C}, or equivalently
to the small amplitude behavior of the quadratic-cubic Swift-Hohenberg
equation (SH23) as described, for instance, in Refs.~\citep{BurkeK:2006,Knobloch:2015de}.
Specifically, on an infinite domain one expects that the primary instability
to the wrinkled state is accompanied by the simultaneous appearance
of \textit{two} branches of spatially localized solutions, one characterized
by deformation that peaks in the center of the domain and the other
that dips in the center. The theory tells us that on an infinite domain
these states are described by the normal form for the reversible Hopf
bifurcation (in space) with 1:1 spatial resonance \citep{Iooss}.
In particular, the branches of localized solutions are initially exponentially
close to one another \citep{KZ1,KZ2,Gelfreich2011}, but gradually
separate as the compression of the sheet increases and the solutions
begin to localize.

Of course, in the experiments the sheet is of finite length. In this
case existing theory shows that the 1:1 reversible Hopf bifurcation
breaks apart into a primary bifurcation to a wrinkled state followed,
at small amplitude, by a secondary bifurcation that creates a pair
of spatially modulated wrinkled states, one with maxima in the center
and the other with minima in the center. This bifurcation is an example
of an Eckhaus instability of the periodic wrinkled state and its dependence
on the domain length has been studied in detail \citep{Bergeon:2008fi}.
In particular, it is known that subsequent secondary bifurcations
lead to the so-called multipulse states, i.e., states with more than
one localized structure within the available domain \citep{BurkeK:2009}.
However, since the periodically wrinkled state does not exhibit a
saddle-node bifurcation the correspondence with SH23 does not extend
to larger amplitude states; in particular, homoclinic snaking \citep{BurkeK:2006,Knobloch:2015de}
is absent.

In this paper we use the similarity between the present problem and
the Swift-Hohenberg equation to identify new classes of localized
wrinkled structures on finite domains \citep{Peletier}. We focus
on the case of a thin elastic sheet allowing us to neglect its contribution
to the gravitational potential energy when its is deformed. In this
case the model equation is odd in the dependent variable, a symmetry
respected by the hinged boundary conditions we employ. As a consequence
the system admits four types of localized states instead of the two
mentioned above. two even parity states under spatial reflection and
two odd parity states. Of these the even states correspond to deformation
that either peaks or dips in the center of the domain while the odd
parity states consist of symmetry-related peak-trough combinations.
Consequently the system exhibits \textit{four} branches of spatially
localized states, and so its properties in fact resemble more closely
those of the small amplitude states in the cubic-quintic Swift-Hohenberg
equation \citep{BurkeK:2007a,BurkeK:2007b} rather than those of the
quadratic-cubic Swift-Hohenberg equation.

Each of these solutions is a single fold solution, with a fold either
in the center of the sheet or split between the two end points. However,
the problem also admits a vast array of multifold states consisting
of both identical and nonidentical folds at different locations along
the sheet. We compute here both single fold and multifold states using
numerical continuation techniques that have proved invaluable in the
studies of the Swift-Hohenberg equation, as well as weakly nonlinear
theory of the type first employed in the present context in \citep{Oshri:2015bi}
and discuss the stability properties of these distinct states.

The paper is organized as follows. In the following subsection we
summarize the basic properties of the model we use. We then focus
on finite sheets, and discuss in turn periodic structures in sheets
with hinged ends and the wrinkle-to-fold transition. The word 'fold'
is here employed to mean a strongly spatially localized deformation.
We compute fold solutions with both positive and negative deformation
in the center, as well as multifold structures. We also compute the
free energy of these structures, and identify structures corresponding
to minima of the free energy with the stable states observed in experiments.
Our results can be considered to be an extension of the work of \cite{2011PhRvL.107p4302D,Rivetti:2013kk}
to finite domains and extend the analysis in \cite{Oshri:2015bi,Marple:2015kq}
to multifold states.

\subsection{Governing equations \label{subsec:Governing-equations}}

\begin{figure}
\begin{centering}
\includegraphics{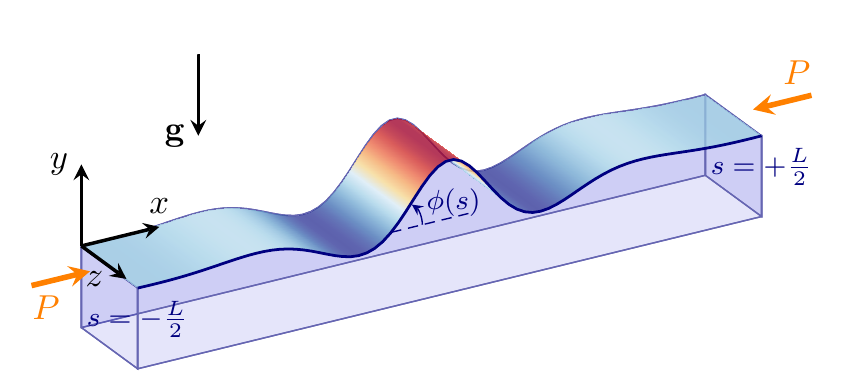} 
\par\end{centering}
\caption{(Color online) Schematic view of the formation of a wrinkled pattern
on an elastic sheet over a fluid substrate under compression. The
shape of the sheet can be parametrized in terms of the local angle
$\phi(s)$ that the sheet forms with the horizontal. The parameter
$s$ represents the arclength along the sheet, whose total length
$L$ is fixed.\label{fig:scheme}}
\end{figure}

Consider an elastic sheet of length $L$ and bending modulus $B$
lying on a semi-infinite fluid substrate of density $\rho$. In static
equilibrium under gravity, a compression $\Delta$ in the $x$ direction
will deform the sheet in the $y$ direction as shown in Fig.~\ref{fig:scheme}.
The shape of the sheet can be parametrized in terms of the arclength
$s$ and the local angle $\phi(s)$ between the sheet and the horizontal
axis. Cartesian coordinates $\left[x(s),y(s)\right]$ are related
to this parametrization through $x(s)=\int_{-\frac{L}{2}}^{s}\cos\phi(s)\,\mbox{d}s,\text{\quad}y(s)=\int_{-\frac{L}{2}}^{s}\sin\phi(s)\,\mbox{d}s.$
Accordingly, the total energy stored (per unit of length in the $z$
direction) due to the deformation is 
\begin{equation}
E\left[\phi\right]=\int_{-L/2}^{+L/2}\mbox{d}s\,\left(\frac{1}{2}B\dot{\phi}^{2}+\frac{1}{2}\rho gy^{2}\cos\phi\right),\label{eq:total_energy}
\end{equation}
where the dot denotes derivative with respect to $s$ and $g$ is
the acceleration due to gravity. The first term is the bending energy
while the second represents the gravitational potential energy of
the deformed state. The sheet shape is linked to the total compression
through 
\begin{equation}
\Delta\left[\phi\right]=\int_{-L/2}^{+L/2}\mbox{d}s\,\left(1-\cos\phi\right).\label{eq:compression}
\end{equation}
Thus the load $P$ required to keep the system in equilibrium is equal
to $dE/d\Delta$.

The natural lengthscale of the system, $\lambda=\left[B/(\rho g)\right]^{1/4}$,
can be used to nondimensionalize the quantities $x,y,s,\Delta$ and
$L$. Likewise, the energy $E$ can be scaled by $B/\lambda$ and
$P$ by $B/\lambda^{2}$. This rescaling leads to a simpler version
of Eq.~\eqref{eq:total_energy} with $B=\rho g=1$.

We study quasistatic deformations of an elastic sheet under the following
two conditions: (i) a fixed load $P$ (\emph{dead loading}), or (ii)
fixed compression $\Delta$ (\emph{rigid loading}). An equation for
the shape of the sheet, $\phi(s)$, can be derived in both cases using
a Lagrangian approach as shown in \citep{Audoly:2011cq,2011PhRvL.107p4302D}.
Whether minimizing the functional $E[\phi]$, the energy, for given
$\Delta$ (rigid loading) or, equivalently, minimizing the functional
$G\left[\phi\right]\equiv E-P\Delta$, the free energy, for given
$P$ (dead loading), one finds that $\phi(s)$ satisfies 
\begin{equation}
\ddddot{\phi}+\frac{3}{2}\dot{\phi}^{2}\ddot{\phi}+p\ddot{\phi}+\sin\phi=0,\label{eq:main}
\end{equation}
where $p$ depends on the boundary conditions. We adopt here hinged
boundary conditions, i.e., the conditions $\left.y\right|_{\pm L/2}=\left.\ddot{y}\right|_{\pm L/2}=0$
(or, equivalently, $\left.\dot{\phi}\right|_{\pm L/2}=\left.\dddot{\phi}\right|_{\pm L/2}=0$),
so $p$ is related to the static load $P$ by \citep{Oshri:2015bi}
\begin{equation}
P=p\cos\phi\left(\pm L/2\right)-\ddot{\phi}\left(\pm L/2\right)\sin\phi\left(\pm L/2\right).\label{eq:Ptilde}
\end{equation}
With these boundary conditions the system is invariant under the transformations
$\phi\rightarrow-\phi$ and $s\rightarrow-s$, which are linked to
invariance with respect to $x\rightarrow-x$ and $y\rightarrow-y$.

Equation~\eqref{eq:main} conserves the following quantity along
the sheet: 
\[
H\left[\phi\right]=\dddot{\phi}\mbox{\ensuremath{\dot{\phi}}}-\frac{1}{2}\ensuremath{\ddot{\phi}}^{2}+\frac{3}{8}\dot{\phi}^{4}+\frac{1}{2}p\dot{\phi}^{2}-\cos\phi.
\]
For hinged sheets $H=\left.-\ensuremath{\frac{1}{2}\ddot{\phi}^{2}}\right|_{\pm L/2}-\cos\left.\phi\right|_{\pm L/2}$.
Throughout this article both dead and rigid loading approaches will
be considered and the results will therefore be represented either
in terms of the quantities $\left(P,G\right)$ or $\left(\Delta,E\right)$,
as appropriate. The difference between these two approaches is fundamental
importance for the stability analysis as discussed in Sec.~\ref{sub:Discussion}.

\subsection{Exact solutions for an infinite sheet}

Despite the presence of nonlinear terms, Eq.~\eqref{eq:main} admits
explicit solutions for the case of an infinite sheet. Two families
of spatially \textit{localized} solutions can be found \citep{2011PhRvL.107p4302D}:
a family of symmetric solutions, 
\begin{equation}
\phi_{s}\left(s\right)=4\arctan\left[\frac{\kappa\sin k\left(s-s_{0}\right)}{k\cosh\kappa\left(s-s_{0}\right)}\right]\label{eq:symmetric}
\end{equation}
and a family of antisymmetric solutions: 
\begin{equation}
\phi_{a}\left(s\right)=4\arctan\left[\frac{\kappa\text{\ensuremath{\cos}}k\left(s-s_{0}\right)}{k\cosh\kappa\left(s-s_{0}\right)}\right].\label{eq:anti-symmetric}
\end{equation}
Here $s=s_{0}$ corresponds to the symmetry point while $k=\frac{1}{2}\sqrt{2+P}$
and $\kappa=\frac{1}{2}\sqrt{2-P}$, i.e., the solutions exist only
for $P<2$. Evaluation in the limit $s\rightarrow\pm\infty$ shows
that for this case $P=p$. The extra parameter $s_{0}$ is a result
of the invariance of the infinite sheet under transformations of the
form $s\rightarrow s+{\rm const}$. In the literature, localized solutions
are referred to as \emph{folds}. These solutions are members of a
wider family of solutions, obtained by adding an extra phase to the
trigonometric function, i.e., 
\begin{equation}
\phi_{f}\left(s\right)=4\arctan\left[\frac{\kappa\text{\ensuremath{\cos}}\left[k\left(s-s_{0}\right)-\varphi_{f}\right]}{k\cosh\kappa\left(s-s_{0}\right)}\right].\label{eq:general}
\end{equation}
Thus the symmetric and antisymmetric solutions correspond to the particular
cases $\varphi_{f}=\pi/2$ and $\varphi_{f}=0$, respectively. The
asymmetric family ($\varphi_{f}\neq0,\pi/2$) has been shown to describe
the shape of an elastic sheet when extracted from a fluid bath \citep{Rivetti:2013kk}.

On the other hand, periodic solutions on an infinite sheet can be
found in terms of Jacobi elliptic functions, 
\begin{equation}
\phi\left(s\right)=2\arcsin\left[k\sn_{k}\left(qs-\vartheta_{0}\right)\right],\label{eq:periodic_phi}
\end{equation}
or, equivalently, 
\begin{equation}
y\left(s\right)=\frac{2k}{q}\cn_{k}\left(qs-\vartheta_{0}\right),\label{eq:periodic_h}
\end{equation}
and exist for $0<k<1$. The quantity $q$ is related to $k$ and the
static load $P$ by the implicit relation 
\begin{equation}
P=q^{2}+\left(1-2k^{2}\right)q^{-2}.\label{eq:periodic_P}
\end{equation}
Thus $q^{2}-q^{-2}<P<q^{2}+q^{-2}$. The solutions \eqref{eq:periodic_phi}
have a period equal to $4\K(k)/q$, where $\K\left(k\right)$ is the
complete elliptic integral of the first kind. Periodic solutions are
usually referred to as \emph{wrinkles} in the literature, cf. \citep{Cerda:2003go,Brau:2013jn}
and references therein.

\subsection{Explicit periodic solutions for finite sheets\label{sub:Periodic}}

Exact periodic solutions of Eq.~\eqref{eq:main} for finite sheets
can easily be found by matching properly both $\vartheta_{0}$ and
the period of Eq.~\eqref{eq:periodic_phi} with the boundary conditions.
For hinged boundary conditions, it follows that 
\begin{equation}
\vartheta_{0}=\nu_{n}\K(k),\quad q_{n}=\frac{2n\K(k)}{L},\quad n=1,2,3...,\label{eq:periodic_finite}
\end{equation}
where $\nu_{n}=1$ for $n$ odd and $\nu_{n}=0$ for $n$ even (see
\citep{Oshri:2015bi} for details). As expected, the finiteness of
the sheet discretizes the parameter $q$ into discrete families $q_{n}(k)$
labeled by the index $n$ (the number of half-periods of the solution).
Solutions of this type will be referred to as $\phi_{n}$. Here $k$
quantifies the amplitude of the solution as measured, for example,
by the mean compression per unit length $\overline{\Delta}\equiv\Delta/L$:
\begin{equation}
\overline{\Delta}=2I(k).\label{eq:periodic_delta}
\end{equation}
Here $I(k)\equiv1-\E(k)/\K(k)$ and $\E(k)$ is the complete elliptic
integral of the second kind. Remarkably, this expression does not
depend on $n$.

\begin{figure}
\begin{centering}
\includegraphics{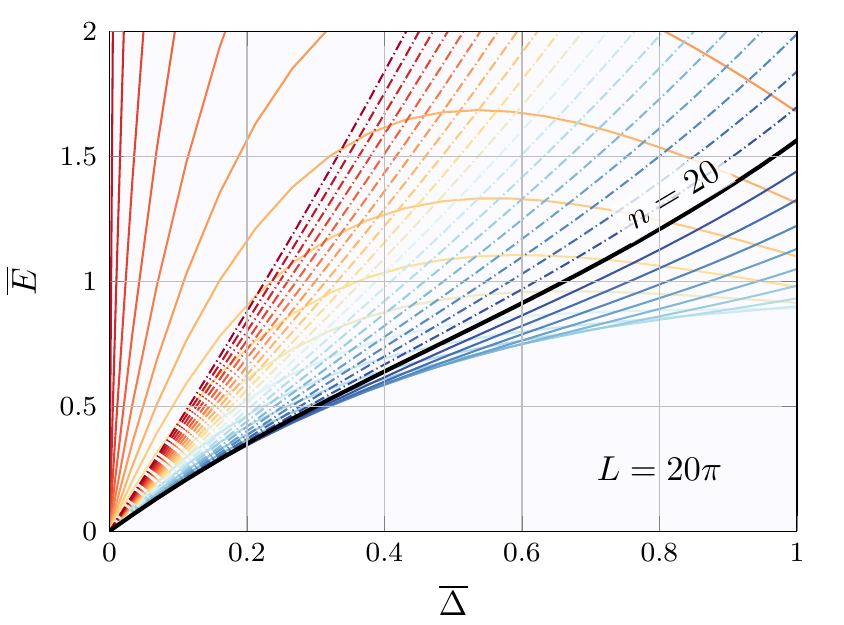} 
\par\end{centering}
\caption{(Color online) Branches of periodic states $\phi_{n}$ in the $(\overline{\Delta},\overline{E})$
plane for a sheet of fixed length $L=20\pi$. Each curve corresponds
to a different family $q_{n}(k)$, $n=1,2,...,40$, with $k$ related
to $\overline{\Delta}$ by Eq.~\eqref{eq:periodic_delta}. The branch
for $n=20$ has the least energy near the common bifurcation point
at the origin. Dash-dotted branches correspond to $n>20$.\label{fig:periodic}}
\end{figure}

The mean energy per unit length, $\overline{E}\equiv E/L$, is given
by 
\begin{equation}
\overline{E}=2\left(q_{n}^{2}+q_{n}^{-2}\right)\left[k^{2}-I(k)\right]-\left[a_{k}-b_{k}I(k)\right],\label{eq:periodic_energy}
\end{equation}
where $a_{k}=8q^{-2}\left(k^{4}+k^{2}\right)/15$ and $b_{k}=16q^{-2}\left(k^{4}-k^{2}+1\right)/15.$
The mean free energy $\overline{G}=\overline{E}-P\overline{\Delta}$
now follows from Eqs.~\eqref{eq:periodic_delta}–\eqref{eq:periodic_energy}.
For a given $n$ and $L$, Eqs.~\eqref{eq:periodic_P}–\eqref{eq:periodic_energy}
determine $\left(P,\overline{\Delta},\overline{E},\overline{G}\right)$
in terms of the single continuous parameter $k$. Figure~\ref{fig:periodic}
depicts the resulting solution branches in the $\left(\overline{\Delta},\overline{E}\right)$
plane for $L=20\pi$ and different $n$ values.

Periodic solutions for finite sheets bifurcate from the trivial state
$\phi(s)=0$ at $k=0$. Since $E(0)=K(0)=\pi/2$, each periodic branch
$\phi_{n}$ bifurcates from the $P$-axis at 
\begin{equation}
P_{n}^{*}=q_{n}^{2}+q_{n}^{-2}\label{eq:periodic_p*}
\end{equation}
and $\Delta_{n}^{*}=0$. Thus $P_{n}^{*}\geq2$ with equality at $n=L/\pi$.
To determine which solution emerges spontaneously when the sheet is
compressed, we expand Eqs.~\eqref{eq:periodic_delta} and \eqref{eq:periodic_energy}
around $k=0$, yielding $\overline{E}\approx\left(q^{2}+q^{-2}\right)k^{2}$
and $\overline{\Delta}\left(k\right)\approx k^{2}.$ It follows that
$\overline{E}\approx\left(q^{2}+q^{-2}\right)\overline{\Delta}.$
The branch with minimal energy in the $\left(\overline{\Delta},\overline{E}\right)$
plane near the trivial state is that which minimizes the slope $q_{n}^{2}+q_{n}^{-2}\approx P_{n}^{*}$
of the energy. Hence, the branch bifurcating at the lowest $P_{n}^{*}$
value is the only one that is stable in the vicinity of the primary
bifurcation.

The branches $\phi_{n}$ and $\phi_{n+1}$ bifurcate simultaneously
from the trivial state whenever $L_{(n,n+1)}=\pi\sqrt{n(n+1)}$ and
do so at $P_{(n,n+1)}^{*}=2+\left[n(n+1)\right]^{-1}$. These conditions
define the range of $L$ for which the $n$th solution is stable:
$L_{(n-1,n)}<L<L_{(n,n+1)}$. As $L\rightarrow\infty$, the width
of this range converges to $\pi$ and its center to $\pi n$.

Far from the primary bifurcation, corresponding to $k=0$, other branches
may display lower energies than the branch that minimizes $P_{n}^{*}$.
This is illustrated in Fig.~\ref{fig:periodic}, which shows that
the lower $n$ branches eventually cross the $n=20$ branch as $\overline{\Delta}$
increases.

\section{Folds in Finite Sheets\label{sec:Folds}}

\subsubsection{Numerical continuation of localized patterns\label{sub:Numerical-continuation}}

Since no explicit exact localized solutions of Eq.~\eqref{eq:main}
exist for elastic sheets of finite length and hinged boundary conditions,
we employ numerical continuation \citep{1991IJBC....1..493D,1991IJBC....1..745D}
to follow different solution types of this problem through parameter
space. For this purpose, we implemented Eq.~\eqref{eq:main} in the
AUTO$^{\text{©}}$ code \citep{doedel08auto-07p} with hinged boundary
conditions, with the independent variable $s$ transformed so the
original half domain $\left[-L/2,0\right]$ becomes $\left[0,1\right]$.
Since the quantities $\left(\overline{\Delta},\overline{E},\overline{G},P\right)$
are all functions of the parameters $\left(p,L\right)$ of the system,
we plot the results as a function of $P$ instead of $p$, for fixed
length $L=l\pi$ ($l\in\mathbb{R}$).

\begin{figure}
\begin{centering}
\includegraphics{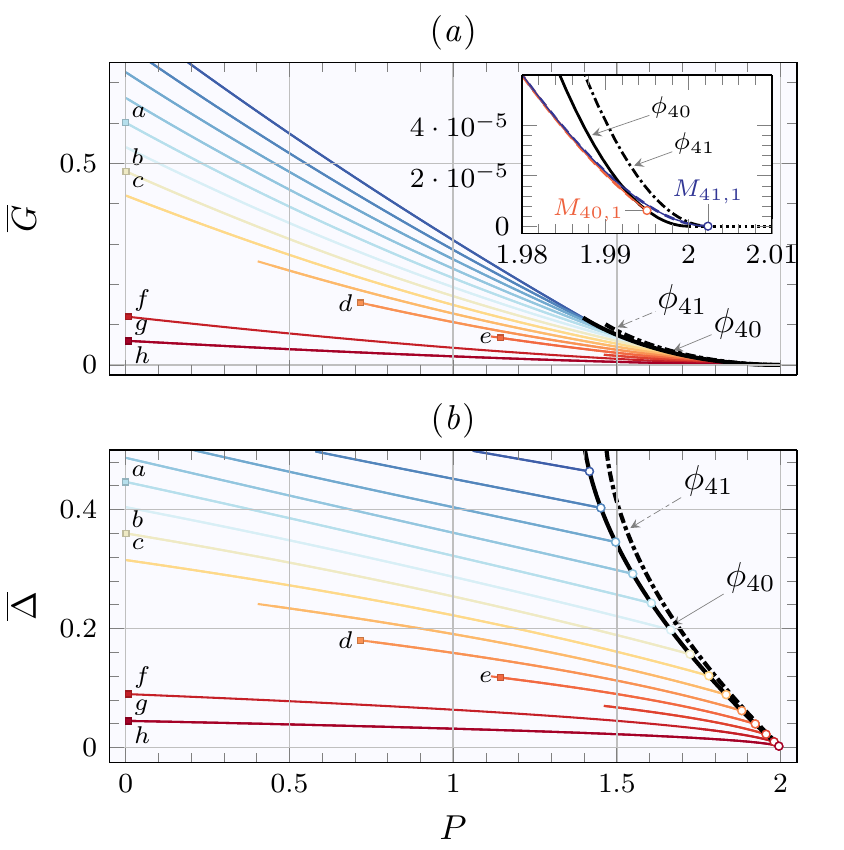} 
\par\end{centering}
\caption{(Color online) Results from numerical continuation of Eq.~\eqref{eq:main}
subject to hinged boundary conditions, showing different solution
branches in the form (\textit{a}) $\overline{G}(P)$ and (\textit{b})
$\overline{\Delta}(P)$, both for $L=40\pi$. The thick black line
represents the periodic state $n=40$, which undergoes several bifurcations
as $P$ decreases (see text for details). The other branches represent
multifold states. The solutions shown in Fig.~\ref{fig:nc-sol} correspond
to the labeled symbols{\tiny{}{}{}{} $\blacksquare$} (\textit{b}-\textit{c}
and \textit{g}-\textit{h} overlap). The inset in (\textit{a}) shows
the region near the origin where the symmetric and antisymmetric fold
solutions first appear. Bifurcation points are marked as {\small{}{}{}{}$\circ$}.
The $n=41$ periodic branch (dashed line) is also displayed. No discernible
difference between the results of the analysis of Sec.~\ref{sub:Approximation}
(thin lines) and the numerical continuation results is seen.}
\label{fig:nc} 
\end{figure}

Numerical continuation requires an initial guess. Taking advantage
of the existence of explicit periodic solutions, we performed continuation
on a $128$ point spatial grid starting from solutions \eqref{eq:periodic_phi}
for a given $n$ and a high value of $k$, i.e., $\phi_{n}$ far from
the trivial state. The code was tuned to detect bifurcation points
along this branch as $P$ increases towards the onset value $P^{*}$.
Figure~\ref{fig:nc}(\textit{a}-\textit{b}) displays the branches
$\phi_{40}$ and $\phi_{41}$ in black, both for $L=40\pi$. Each
periodic branch displays a set of bifurcation points with decreasing
separation as $P$ increases towards $P^{*}$. For $\phi_{40}$ these
bifurcation points are shown Fig.~\ref{fig:nc}(\textit{b}). These
bifurcation points will be referred to as $M_{n,m}$, where the integer
$n$ specifies the primary periodic state and the integer $m$ counts
the bifurcations from this state starting from the primary bifurcation
$P_{n}^{*}$. Thus $M_{40,1}$ denotes the secondary bifurcation closest
to $P_{40}^{*}$.

\begin{figure*}
\begin{centering}
\includegraphics{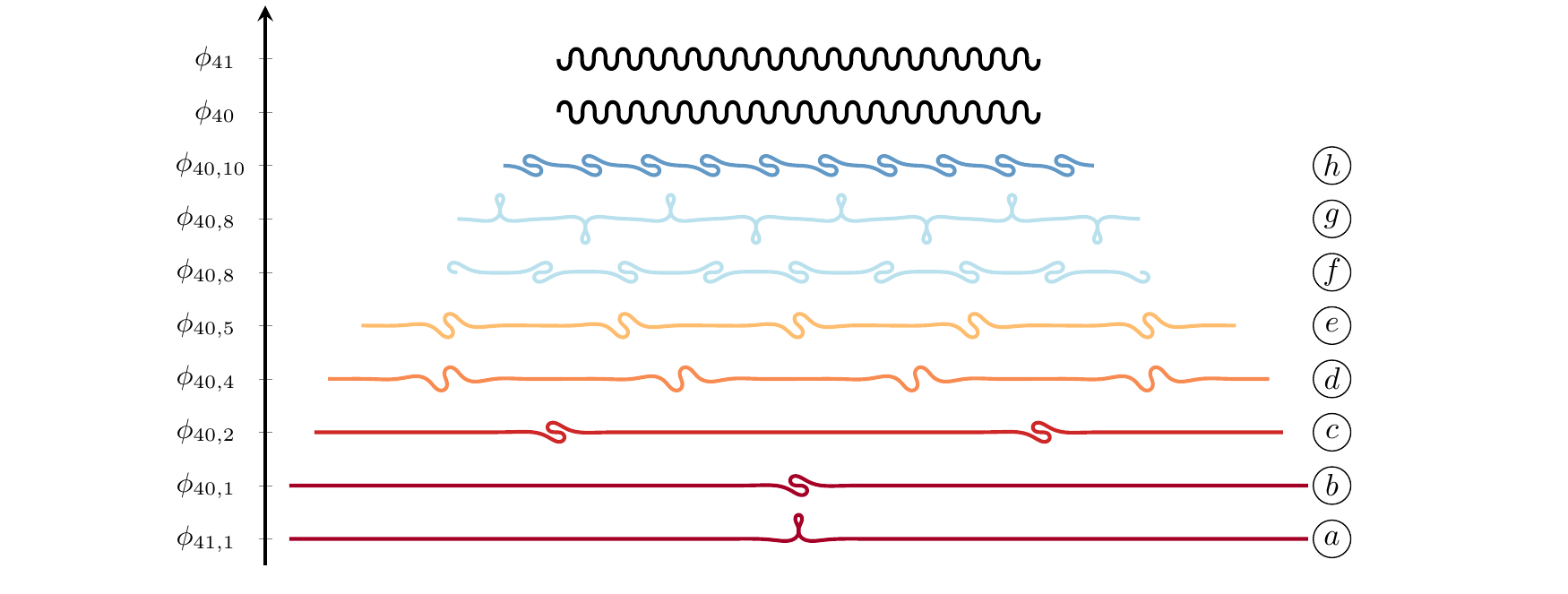} 
\par\end{centering}
\caption{(Color online) Antisymmetric solutions ($b$-$h$) from numerical
continuation for different values of $m$ defined in the text and
$n=40$ computed on a domain of length $L=40\pi$. The location of
each solution in parameter space is indicated in Fig.~\ref{fig:nc}.
The solutions acquire additional folds as $m$ increases. The symmetric
solution $\phi_{41,1}$ (\textit{a}) bifurcates from the $n=41$ periodic
state, and does so very close to the trivial state, as shown in the
inset in Fig.~\ref{fig:nc}(\textit{a}). The periodic solutions $\phi_{40}$
and $\phi_{41}$ at high compression ($\overline{\Delta}=0.55$) are
included for reference.\label{fig:nc-sol}}
\end{figure*}

At each of the secondary bifurcation points $M_{n,m}$ we switched
the branch direction and ran the numerical continuation for decreasing
$P$ until $P=0$ was attained. The new branches are also displayed
in Fig.~\ref{fig:nc} but now in color. Some of the solutions far
from the bifurcation points $M_{n,m}$ are depicted in Fig.~\ref{fig:nc-sol}.
A simple inspection {[}see inset in Fig.~\ref{fig:nc}(\textit{a}){]}
shows that the branch starting at $M_{40,1}$ corresponds to a localized
antisymmetric fold while that starting at $M_{41,1}$ is a symmetric
fold. The former bifurcation point is located quite far from $P_{40}^{*}$,
i.e., at finite amplitude, while the latter lies very close to $P_{41}^{*}$.
At this value the $\phi_{41}$ state barely departs from the trivial
state. For $L=41\pi$ the situation is the opposite: the symmetric
fold is the one that bifurcates at finite amplitude (from the $\phi_{41}$
branch), while the antisymmetric fold bifurcates at a very small amplitude
(from the $\phi_{42}$ branch). In any case, the symmetric and antisymmetric
branches approach each other very fast as they depart from their respective
branch point with decreasing $P$. 

Solutions $\phi_{n,m}$ with $m>1$ represent $m$-fold solutions,
i.e., arrays of $m$ equispaced folds. Examples are shown in Fig.~\ref{fig:nc-sol}.
Although reported in \cite{Marple:2015kq}, these multifold solutions
have not been described in detail. Multifold solutions $\phi_{n,m}$
for a fixed $l$ do not exist for arbitrary $n$ and $m$. For instance,
for $l=40,$ one-fold solutions ($m=1$) are observed only for $n=40$
and $41$ while two-fold solutions ($m=2$) are only found for $n=39,40,41,42$.
In general, $2m+2$ solutions with $m$ folds were detected for $n$
varying from $l-m+1$ to $l+m$.

\subsection{Weakly modulated solutions\label{sub:Approximation}}

Multiscale analysis describes successfully several features of this
system and, in particular, the wrinkle-to-fold transition \citep{Brau:2010fia,Audoly:2011cq}.
Herein, we use this approach to understand the emergence of the solutions
obtained in Sec.~\ref{sub:Numerical-continuation}. The system is
assumed to be subject to hinged boundary conditions and finite. The
domain of $s$ is mapped into $s\in\left[0,L\right]$ for the sake
of simplicity in the application of the boundary conditions.

We seek multiscale solutions of Eq.~\eqref{eq:main} in the form
\begin{equation}
y\left(s\right)=\sum_{j=1}^{\infty}\epsilon^{j}u_{j}\left(s,S=\epsilon s\right),\label{eq:multi-scale}
\end{equation}
where the functions $u_{j}$ are of $\mathcal{O}\left(1\right)$ and
$\epsilon$ is a small parameter measuring the distance from some
primary instability threshold, $\epsilon^{2}\equiv p^{*}-p$. 
Since $\phi\left(s\right)=\arcsin(\partial_{s}y)$ with $\partial_{s}$
now replaced by $\partial_{s}+\epsilon\partial_{S}$, Eq.~\eqref{eq:main}
generates a hierarchy of equations for the functions $u_{j}\left(s,S\right)$.
The first three equations are 
\begin{eqnarray}
\mathcal{O}\left(\epsilon^{1}\right):\quad\mathcal{L}_{p^{*}}\partial_{s}u_{1} & = & 0,\nonumber \\
\mathcal{O}\left(\epsilon^{2}\right):\quad\mathcal{L}_{p^{*}}\partial_{s}u_{2} & = & -\mathcal{L}'_{p^{*}}\partial_{s}u_{1},\label{eq:hierarchy}\\
\mathcal{O}\left(\epsilon^{3}\right):\quad\mathcal{L}_{p^{*}}\partial_{s}u_{3} & = & -\mathcal{L}'_{p^{*}}\partial_{s}u_{2}-\mathcal{N}_{p^{*}}\partial_{s}u_{1},\nonumber 
\end{eqnarray}
where the fourth-order linear operators $\mathcal{L}_{p^{*}}$ and
$\mathcal{L}'_{p^{*}}$ and the nonlinear operator $\mathcal{N}_{p^{*}}$
are, respectively, defined as 
\begin{eqnarray*}
\mathcal{L}_{p^{*}}w & \equiv & \left(\partial_{s}^{4}+p^{*}\partial_{s}^{2}+1\right)w,\\
\mathcal{L}'_{p^{*}}w & \equiv & \left(5\partial_{s}^{4}+3p^{*}\partial_{s}^{2}+1\right)w,\\
\mathcal{N}_{p^{*}}w & \equiv & \frac{1}{2}\ddddot{w}w^{2}+10\ddot{w}''+\frac{15}{2}\ddot{w}w''+3\ddot{w}^{2}w+\ldots\\
 &  & +4\dddot{w}\dot{w}w-\ddot{w}+p^{*}\left[\frac{1}{2}\ddot{w}w^{2}+3w''+\dot{w}w\right].
\end{eqnarray*}
Here the overdots denote derivatives with respect to $s$ while the
primes denote derivatives with respect to $S$. At the instability
threshold $p^{*}=2$ the linear operators reduce to $\mathcal{L}_{2}\equiv\left(\partial_{s}^{2}+1\right)^{2}$
and $\mathcal{L}_{2}'\equiv\left(\partial_{s}^{2}+1\right)\left(5\partial_{s}^{2}+1\right)$
and have a common kernel spanned by the null eigenfunctions of the
second order linear operator $\partial_{s}^{2}+1$. In this case the
$\mathcal{O}\left(\epsilon^{1}\right)$ solution of \eqref{eq:hierarchy}
is $u_{1}\left(s,S\right)=A\left(S\right)\sin\left(s-\theta\right)$,
i.e., the dominant term $u_{1}(s,S)$ is a carrier wave with wave
number $q=1$ modulated by a slowly varying envelope $A(\epsilon s)$.
It now follows that the right side of the $\mathcal{O}\left(\epsilon^{2}\right)$
equation vanishes, and hence that $u_{2}\equiv0$. An equation for
$A(S)$ is therefore obtained by imposing a solvability condition
at $\mathcal{O}\left(\epsilon^{3}\right)$. A straightforward calculation
yields 
\begin{equation}
A''-\frac{1}{4}A+\frac{1}{8}A^{3}=0.\label{eq:jacobi}
\end{equation}
Equation~\eqref{eq:jacobi} has two solutions given by Jacobi elliptic
functions, 
\begin{eqnarray}
\epsilon A_{\dn}(S) & = & 4\epsilon\kappa_{\dn}\dn_{k}\left(\kappa_{\dn}S-\vartheta_{\dn}\right),\label{eq:A_dn}\\
\epsilon A_{\cn}(S) & = & 4\epsilon\kappa_{\cn}{}^{\frac{1}{2}}k\cn_{k}\left(\kappa_{\cn}S-\vartheta_{\cn}\right),\label{eq:A_cn}
\end{eqnarray}
where $\vartheta$ is an arbitrary phase and 
\begin{eqnarray}
\epsilon^{2}\kappa_{\dn}^{2} & = & \frac{1}{4}\left(2-p\right)\left(2-k^{2}\right)^{-1}\label{eq:epsilon_dn}\\
\epsilon^{2}\kappa_{\cn}^{2} & = & \frac{1}{4}\left(2-p\right)\left(2k^{2}-1\right)^{-1}.\label{eq:epsilon_cn}
\end{eqnarray}
Thus each solution branch is parametrized by $k$, just as for the
periodic case in Sec.~\ref{sub:Periodic}. The final step is to multiply
the expressions for the modulation amplitude $A$ by the carrier wave,
$\sin(qs-\theta)$, and match the result to the hinged boundary conditions.
Note that both the modulation amplitude and the carrier wave are periodic
functions 0f $s$.

\subsubsection{Wave number and phase selection\label{sub:Wave-number}}

Since the wave number $q=1$, the weakly modulated solutions satisfy
hinged boundary conditions only for a simple set of combinations of
$\theta,\vartheta,L$ and $\epsilon\kappa$, displayed in Table~\ref{tab:set}.
This set can be identified on realizing that the chosen boundary conditions
are satisfied if and only if one of the two periodic functions, either
the trigonometric or the elliptic one, has a node (zero) at the boundary
while the other has an antinode (local maximum or minimum) at the
same location. 
\begin{center}
\begin{table}[h]
\begin{centering}
\begin{tabular}[t]{c>{\centering}p{1ex}c>{\centering}p{1ex}c>{\centering}p{1ex}c>{\centering}p{1ex}c}
\noalign{\vskip\doublerulesep}  &  & $\theta$  &  & $\vartheta$  &  & $L$  &  & $\epsilon\kappa L$\tabularnewline
\hline 
\hline 
\noalign{\vskip\doublerulesep}$y_{\dn}\left(s\right)$  &  & $0,\pi$  &  & $0,K\left(k\right)$  &  & \hfill{}$l\pi$  &  & \hfill{}$mK\left(k\right)$\tabularnewline
\noalign{\vskip\doublerulesep}$y_{\cn}\left(s\right)$  &  & $0,\pi$  &  & $\pm K\left(k\right)$  &  & \hfill{}$l\pi$  &  & \hfill{}$2mK\left(k\right)$\tabularnewline
\noalign{\vskip\doublerulesep}  &  &  &  &  &  & $\left(l+\frac{1}{2}\right)\pi$  &  & $2\left(m+\frac{1}{2}\right)K\left(k\right)$\tabularnewline
\noalign{\vskip\doublerulesep}  &  & $\pm\frac{1}{2}\pi$  &  & $0,K\left(k\right)$  &  & \hfill{}$l\pi$  &  & \hfill{}$2mK\left(k\right)$\tabularnewline
\noalign{\vskip\doublerulesep}  &  &  &  &  &  & $\left(l+\frac{1}{2}\right)\pi$  &  & $2\left(m+\frac{1}{2}\right)K\left(k\right)$\tabularnewline
\hline 
\end{tabular}
\par\end{centering}
\caption{Set of combinations of $\theta,\vartheta,L$ and $\epsilon\kappa$
that fulfill hinged boundary conditions. Both $l$ and $m$ are positive
integers. The functions $y_{\dn}\left(s\right)$ and $y_{\cn}\left(s\right)$
are given by the product of $\sin\left(s-\theta\right)$ and $\epsilon A_{\dn}$
and $\epsilon A_{\cn}$, respectively. \label{tab:set}}
\end{table}
\par\end{center}

Since both elliptic functions localize as $k\rightarrow1$, some of
the solutions from Table~\ref{tab:set} may display asymmetric folds
at the boundaries. This happens whenever the trigonometric function
has a node at the boundary. These types of solutions were also detected
in our numerical continuation. However, in the remainder of this article,
and for the sake of simplicity, we restrict our analysis to solutions
that do not display folds at the boundaries. The solutions that remain
are summarized in Table~\ref{tab:subset}. Solutions related by the
symmetries $x\rightarrow-x$, $y\rightarrow-y$ have been omitted. 
\begin{center}
\begin{table}[h]
\begin{centering}
\begin{tabular}[t]{c>{\centering}p{1ex}c>{\centering}p{1ex}c>{\centering}p{1ex}c>{\centering}p{1ex}c}
\noalign{\vskip\doublerulesep}  &  & $\theta$  &  & $\vartheta$  &  & $L$  &  & $\epsilon\kappa L$\tabularnewline
\hline 
\hline 
\noalign{\vskip\doublerulesep}$y_{\dn}\left(s\right)$  &  & $0$  &  & $K\left(k\right)$  &  & \hfill{}$l\pi$  &  & \hfill{}$2mK\left(k\right)$\tabularnewline
\noalign{\vskip\doublerulesep}$y_{\cn}\left(s\right)$  &  & $\frac{1}{2}\pi$  &  & $K\left(k\right)$  &  & \hfill{}$l\pi$  &  & \hfill{}$2mK\left(k\right)$\tabularnewline
\hline 
\end{tabular}
\par\end{centering}
\caption{Subset of the combinations of $\theta,\vartheta,L$ and $\epsilon\kappa$
that fulfill the assumed boundary conditions. Solutions related by
the symmetries $x\rightarrow-x$ and $y\rightarrow-y$ have been omitted
as have any solutions that develop folds at the boundaries as $k\rightarrow1$.
\label{tab:subset}}
\end{table}
\par\end{center}

\subsubsection{Branches in parameter space\label{sub:Branches-in-parameter}}

The weakly modulated solutions $y(s)=A(\epsilon s)\sin(s-\theta)$
can be used to build expansions for the static load $P$, the mean
energy $\overline{E},$ the mean compression $\overline{\Delta}$
and the mean free energy $\overline{G}$ for a sheet of finite length
$L$. For a static load $P$ the condition $p=2-\epsilon^{2}$ translates
into $P=2-\epsilon^{2}+\mathcal{O}\left(\epsilon^{4}\right)$ using
Eq.~\eqref{eq:Ptilde} and the solutions obtained in Sec.~\ref{sub:Wave-number}.
Accordingly, $\left(\overline{E},\overline{\Delta},\overline{G}\right)=\sum_{j=1}\epsilon^{j}\left(\overline{E}_{j},\overline{\Delta}_{j},\overline{G}_{j}\right)$.
The expressions for $\left(\overline{E}_{j},\overline{\Delta}_{j},\overline{G}_{j}\right)$
can be simplified by averaging over one period of the fast oscillation,
e.g., $\int_{0}^{L}\mathrm{d}s\,\sin^{2}\left(s\right)A\left(\epsilon s\right)=\frac{1}{2}\int_{0}^{L}\mathrm{d}s\,A\left(\epsilon s\right)$.
As a consequence, the first and third order terms all vanish and the
remaining quantities only depend on the envelope $A(S)$. The following
expressions for the quantities $P$, $\overline{G}$, $\overline{\Delta}$,
$\overline{E}$ to $\mathcal{O}\left(\epsilon^{4}\right)$ are obtained:
\begin{equation}
\begin{array}{rcl}
P & = & {\displaystyle 2-\epsilon^{2}+\mathcal{O}\left(\epsilon^{4}\right),}\\
\overline{G} & = & {\displaystyle \frac{1}{2}\epsilon^{4}\left[\frac{1}{2}\overline{A^{2}}+2\overline{A'^{2}}-\frac{1}{8}\overline{A^{4}}\right]},\\
\overline{\Delta} & = & {\displaystyle \frac{1}{4}\epsilon^{2}\overline{A^{2}}+\frac{1}{4}\epsilon^{4}\left[\frac{3}{16}\overline{A^{4}}+\overline{A'^{2}}\right]},\\
\overline{E} & = & {\displaystyle \frac{1}{2}\epsilon^{2}\overline{A^{2}}+\frac{1}{2}\epsilon^{4}\left[3\overline{A'^{2}}+\frac{1}{16}\overline{A^{4}}\right]}.
\end{array}\label{eq:expansion}
\end{equation}
The overbars indicate spatial averages. After replacing $A$ by $A_{\dn}$
or $A_{\cn}$ from Eqs.~\eqref{eq:A_dn}–\eqref{eq:A_cn}, all the
averaged quantities can be expressed in terms of the complete elliptic
integrals $K(k)$ and $I(k)$ (see Appendix~\ref{sub:Appendix-A}).
For our reduced set of solutions, $\epsilon=2mK(k)/(\kappa L)$, where
$\kappa$ is either $\kappa_{\dn}$ or $\kappa_{\cn}$). Hence, all
solution branches are fully parametrized in terms of $k$. Inversion
of Eqs.~\eqref{eq:epsilon_dn}–\eqref{eq:epsilon_cn} yields an expression
for the static load. For $q=1$ we obtain, through $\mathcal{O}\left(\epsilon^{4}\right)$,
\begin{eqnarray*}
\left[\begin{array}{c}
P_{\dn}\\
P_{\cn}
\end{array}\right] & = & 2-\frac{2^{4}m^{2}}{L^{2}}K^{2}(k)\left[\begin{array}{c}
2-k^{2}\\
2k^{2}-1
\end{array}\right],\\
\left[\begin{array}{c}
\overline{G}_{\dn}\\
\overline{G}_{\cn}
\end{array}\right] & = & \frac{2^{8}m^{4}}{3L^{4}}K^{4}(k)\left[\begin{array}{c}
3-k^{2}-2\left(2-k^{2}\right)I(k)\\
3k^{4}-k^{2}-2\left(2k^{2}-1\right)I(k)
\end{array}\right],\\
\left[\begin{array}{c}
\overline{\Delta}_{\dn}\\
\overline{\Delta}_{\cn}
\end{array}\right] & = & \frac{2^{4}m^{2}}{L^{2}}K^{2}(k)\left[\begin{array}{c}
1-I(k)\\
k^{2}-I(k)
\end{array}\right]+\\
 &  & \frac{2^{6}m^{4}}{3L^{4}}K^{4}(k)\left[\begin{array}{c}
9-2k^{2}-7\left(2-k^{2}\right)I(k)\\
9k^{4}-2k^{2}-7\left(2k^{2}-1\right)I(k)
\end{array}\right],\\
\left[\begin{array}{c}
\overline{E}_{\dn}\\
\overline{E}_{\cn}
\end{array}\right] & = & \frac{2^{5}m^{2}}{L^{2}}K^{2}(k)\left[\begin{array}{c}
1-I(k)\\
k^{2}-I(k)
\end{array}\right]+\\
 &  & \frac{2^{7}m^{4}}{3L^{4}}K^{4}(k)\left[\begin{array}{c}
3+2k^{2}-5\left(2-k^{2}\right)I(k)\\
3k^{4}+2k^{2}-5\left(2k^{2}-1\right)I(k)
\end{array}\right].
\end{eqnarray*}
We use the above expressions to characterize the dnoidal and cnoidal
solutions in a bifurcation diagram, and the parameter $k$ as a measure
of their amplitude. In particular, the condition $k\rightarrow0$
determines the threshold for the appearance of each solution type.

\subsection{Wrinkle-to-fold transition\label{sub:Wrinkle-to-fold}}

\begin{figure}
\begin{centering}
\includegraphics{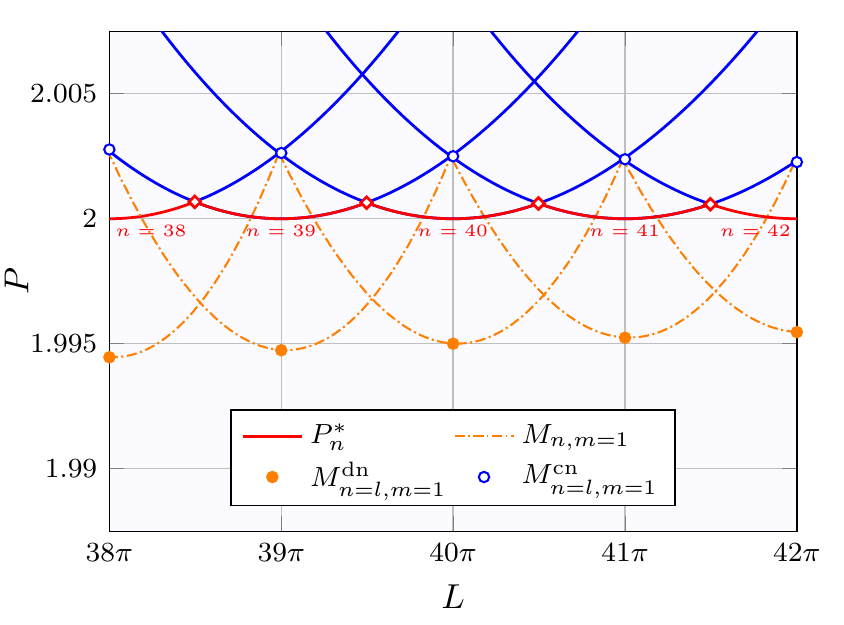} 
\par\end{centering}
\caption{(Color online) Bifurcation thresholds in the $(L,P)$ plane: primary
bifurcation thresholds $P_{n}^{*}$ to periodic states $\phi_{n}$
together with secondary bifurcation thresholds $M_{n,m=1}$ to dnoidal
states (dash-dotted lines) obtained by numerical continuation, each
for $n=38,39,...,42$. The symbol $\diamond$ denotes codimension-two
transitions between $n$ and $n+1$ periodic wrinkle states. The red
filled (resp. blue open) circles indicate the multiscale analysis
predictions of bifurcations to dnoidal (resp. cnoidal) states when
$n=l$ (with $l=L/\pi\in\mathbb{Z}^{+}$). \label{fig:Bifurcation_points}}
\end{figure}

Experiments have shown that an elastic sheet over a fluid substrate
undergoes a wrinkle-to-fold transition when continuously compressed
from its original length \citep{Pocivavsek:2008kz}. The wrinkle-to-fold
transition is characterized by the emergence of a localized fold from
a periodic pattern. Figure~\ref{fig:Bifurcation_points} shows the
critical loads $P_{n}^{*}$ for the appearance of a periodic state
with $n=38,39,...,42$ wavelengths in the domain (solid lines), as
well as the secondary bifurcation thresholds $M_{n=l,m=1}^{\dn}$
for the appearance of the lowest energy dnoidal states ($m=1$) obtained
from the $k\rightarrow0$ limit of the dnoidal solutions of the previous
section (red filled circles). In contrast, the cnoidal solutions bifurcate
from the periodic states within an $\mathcal{O}\left(L^{-3}\right)$
neighborhood of $L=L_{(n-1,n+1)}$, the crossing points for like-parity
modes $n-1$ and $n+1$. The predicted bifurcation thresholds $M_{n=l,m=1}^{\cn}$
(blue open circles) are also obtained from the $k\rightarrow0$ limit
of the corresponding analytical solution. These bifurcation points
and the analytical expressions for the small amplitude solutions present
nearby were used as an input into a numerical continuation routine
and used to extend these thresholds to noninteger values of $L/\pi$
($M_{n,m=1}$ in Fig.~\ref{fig:Bifurcation_points}). Numerical continuation
reveals that the $M_{n,m=1}$ curves resemble parabolas with vertices
located in the vicinity of $M_{n=l,m=1}^{\dn}$. These parabolas extend
between the codimension-two points $L_{(n-2,n)}$ and $L_{(n,n+2)}$
(or approximately, between $M_{n=l-1,m=1}^{\cn}$ and $M_{n=l+1,m=1}^{\cn}$).
Also indicated are the codimension-two points $L=L_{(n,n+1)}$ (open
red diamonds) corresponding to the crossing of periodic modes $n$
and $n+1$, i.e., the points at which the primary bifurcation to a
symmetric wrinkle state changes to an antisymmetric wrinkle state
as $L$ changes. The curves $M_{n,m=1}$ and $P_{n}^{*}$, as well
as the predicted locations $M_{n=l,m=1}^{\dn}$ and $M_{n=l,m=1}^{\cn}$,
are all compressed towards the $P=2$ axis as $L$ increases.

As shown in Fig.~\ref{fig:Bifurcation_points} the multiscale analysis
generates a discrete set of bifurcation points instead of the continuous
curves of bifurcation points obtained by numerical continuation. The
solid curves for $L\in(l\pi,(l+1)\pi)$, $l\in\mathbb{Z}^{+}$, can
only be obtained via numerical continuation. This limitation of the
multiscale analysis has its origin in the structure of the equations
analyzed in Sec.~\ref{sub:Approximation}, which requires $l\in\mathbb{Z}^{+}$
to satisfy simultaneously the boundary conditions and the cancellation
of the right-hand side of the $\mathcal{O}\left(\epsilon^{2}\right)$
equation in Eq.~\eqref{eq:hierarchy}. If we instead follow \citep{Audoly:2011cq,Oshri:2015bi}
and minimize the functional $\overline{G}$ on the assumption that
$y(s)=A(S)\cos(qs)$ we obtain Eq.~\eqref{eq:jacobi} when $q=1$
and a similar equation to Eq. \eqref{eq:jacobi} but with $q$-dependent
coefficients whenever $q>1$. The latter, however, yields inaccurate
solutions, which are an artifact of improper assumption on the $s$-dependence
of $y(s)$ when $q>1$.

\subsection{The $\pi/2$-shifted localized branch}

The exact nonlinear solutions in Eqs.~(\ref{eq:symmetric})–(\ref{eq:anti-symmetric})
show that the symmetric and antisymmetric folds on an infinite sheet
are equivalent from the energy point of view. Such states are therefore
energetically degenerate. However, this is no longer the case on a
finite sheet and the question arises whether whether a sheet whose
wrinkle-to-fold transition yields a symmetric (antisymmetric) fold
may also display an antisymmetric (symmetric) fold. Results from Sec.~\ref{sub:Branches-in-parameter}
provide hints about the dynamical features of these two states. For
$l\in\mathbb{Z}^{+}$ symmetric and antisymmetric single fold states
(i.e., $n=l,m=1$) can be understood in terms of the carrier wave
phase $\theta$. The idea is simple: since both Jacobi elliptic functions
of the approximate solutions in Table~\ref{tab:subset} have a maximum
at the center of the sheet, the choice of whether $\theta=0$ or $\theta=\pi/2$
determines the symmetry of the central fold. For $n=l$ even (odd),
the dnoidal (cnoidal) branch yields the antisymmetric (symmetric)
single fold state and the cnoidal (dnoidal) branch the symmetric (antisymmetric)
state. However, there is an important difference between these two
branches. The cnoidal branch, whose carrier wave is shifted by $\pi/2$
relative to the dnoidal branch, emerges from the trivial state through
a primary bifurcation to the right of the primary bifurcation to the
periodic state (and half as close to it as the dnoidal branch bifurcation
on the left). The resulting predictions of the theory of Sec.~\ref{sub:Branches-in-parameter}
are displayed in the inset of Fig.~\ref{fig:nc}(\textit{a}) and
compared there with the corresponding results from numerical continuation.
The accuracy of the approach is very good.

The theoretical predictions show that in the limit $L\rightarrow\infty$
both branches merge into a single one in terms of $\left(\Delta,E,G,P\right)$.
As $L$ increases, the secondary bifurcation to dnoidal states is
pushed downward along the primary periodic branch while the cnoidal
bifurcation point is also pushed towards the primary bifurcation point.
These two bifurcation points coalesce in the limit $L\rightarrow\infty$.
The corresponding large amplitude dnoidal and cnoidal branches become
rapidly very close as $P$ decreases, a prediction that is confirmed
by numerical continuation to values of $\epsilon$ far beyond $\epsilon=1$,
i.e., $P\ll1$.

\subsection{Multifold solutions}

\begin{figure*}
\begin{centering}
\includegraphics{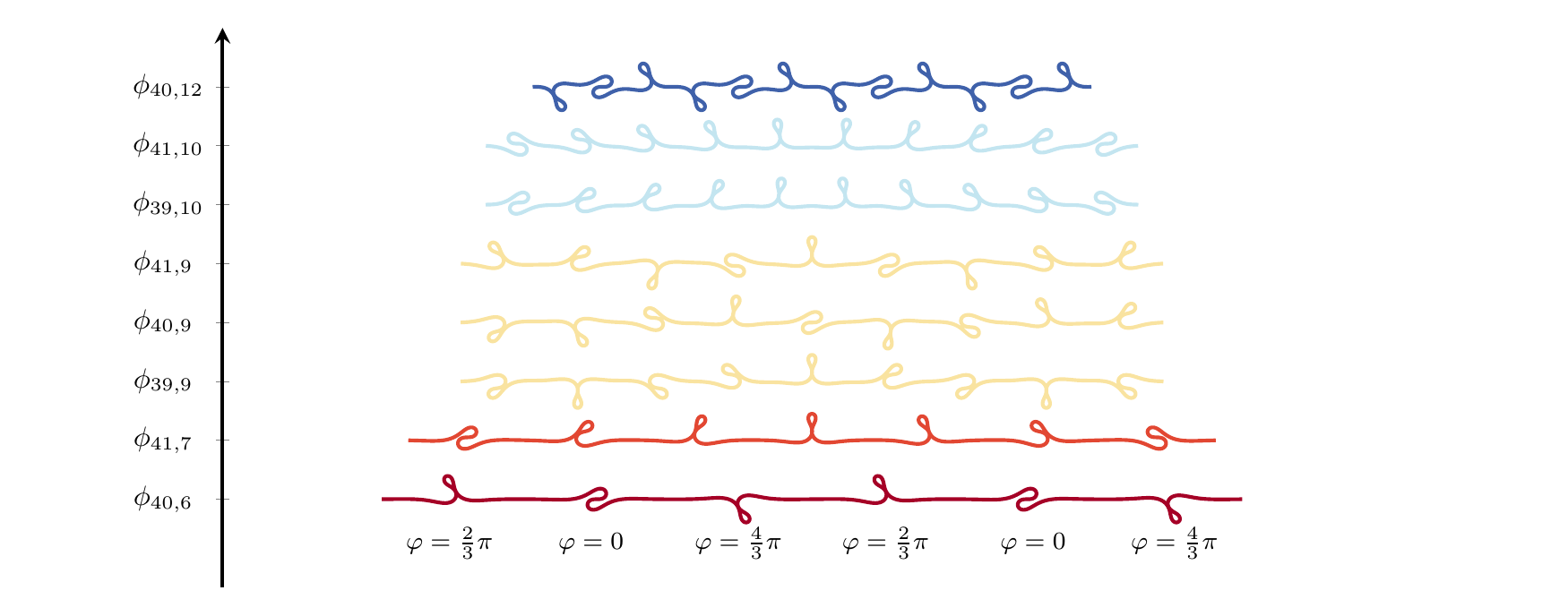} 
\par\end{centering}
\caption{(Color online) A set of spatially modulated periodic solutions from
numerical continuation for a sheet of length $L=40\pi$ and different
values of $n$ and $m$. In all cases $P=1$. The solutions are composed
of distorted folds. While $m$ determines the number of folds, $n$
fixes the shape of each fold (see text for details). For the lowest
state, the local phase of the carrier wave, $\varphi_{f}$, is indicated.
\label{fig:nc-sol-distorted}}
\end{figure*}

The modulation analysis carried out above also provides a clearer
picture of the emergence of multifold ($m\ge2$) solutions for $l\equiv(L/\pi)\in\mathbb{Z}^{+}$.
The condition $q=1$ requires that $n=l$, where the labels $n$ and
$m$, introduced in Sec.~\ref{sub:Numerical-continuation}, represent
the carrier and envelope wave numbers, respectively. The bifurcations
for $m\ge2$ follow the same principle as the branches of localized
states when $m=1$, i.e., the dnoidal $m$-branch bifurcates from
a periodic state $\phi_{n}$ at finite amplitude while the corresponding
cnoidal $m$-branch bifurcates from the periodic state $\phi_{n+m}$
at a very small amplitude, and to the right of the primary bifurcation
to $\phi_{n}$ and half as close to it as the bifurcation to the dnoidal
branch. For example, when $m=2$ the cnoidal branch in general bifurcates
from $\phi_{42}$ and does so very close to the point $P_{42}^{*}$,
although it may bifurcate directly from the trivial state but only
at the crossing point of $P_{38}^{*}$ and $P_{42}^{*}$. Likewise,
the bifurcation points to the $m\ge2$ dnoidal branches spread out
from the primary bifurcation point to the $n$th periodic branch as
$m^{2}$. We also used numerical continuation to follow the $M_{n,m}$
bifurcation points as a function of $L$ as done in Sec.~\ref{sub:Wrinkle-to-fold}
for $M_{n,1}$. Just as the single-fold cnoidal bifurcation point
lies in the vicinity of $L_{\left(n-1,n+1\right)}$, the crossing
point of the primary bifurcations to the $n-1$ and $n+1$ periodic
branches, the $m$-fold cnoidal bifurcation point is located in the
vicinity of $L_{\left(n-m,n+m\right)}$, the crossing point of the
primary bifurcations to the $n-m$ and $n+m$ periodic branches. Likewise,
the $M_{n,m}$ bifurcation curves for noninteger $L/\pi$ resemble
parabolas with vertices near $M_{l=n,m}^{\dn}$ extending from $L_{(n-2m,n)}$
to $L_{(n,n+2m)}$. The range of existence of the parabola clarifies
why the number of $m$-fold states observed is $2m+2$ and which $n$
values are allowed for a given $m$ and $L$.

The analysis also reveals another interesting feature. The choice
of $l$ and $m$ determines (a) if the solution consists of an array
of symmetric or antisymmetric folds, and (\textit{b}) the local shape
of each fold. The rule for (a) is very simple: these states are allowed
if $l$ is divisible by $m$. For the dnoidal solutions, if $l/m$
is even, the solutions will be an array of antisymmetric folds whereas
for $l/m$ odd the solution will be an array of symmetric folds with
alternating orientation. If folds are allowed at both boundaries,
another possible solution, consisting of alternating antisymmetric
folds for $l/m$ even, can be observed. For $l=40$, periodic multifold
states are observed for $m=2,4,5,8,10,20$, as displayed in Fig.~\ref{fig:nc-sol}
($m=20$ is not shown). The figure also shows a state with folds at
either boundary, for $m=8$, to illustrate the complementary family
of periodic solutions.

When $l$ is not divisible by $m$ the folds forming the array become
distorted and resemble the family of arbitrary phase folds described
in \citep{Rivetti:2013kk} for an infinite length sheet. The rule
(\textit{b}) for the local shape of each fold can be understood in
terms of the phase $\varphi_{f}$ in Eq.~\eqref{eq:general}, which
is now fixed by the carrier wave. The $j$th fold of the dnoidal solutions
displays a local phase $\varphi_{f}=-l\pi\left(j-1/2\right)/m$, whereas
for the cnoidal solutions $\varphi_{f}=-\pi/2-l\pi\left(j-1/2\right)/m$.
Accordingly, the fundamental period of an array of folds in a dnoidal
solution is given by $2L/\mbox{GCD}\left(l,2m\right)$ while the fundamental
period for cnoidal solutions is $L/\mbox{GCD}\left(l,m\right)$ provided
$l$ is divisible by $m$ and $l/\mbox{GCD}\left(l,m\right)$ is odd,
or $2L/\mbox{GCD}\left(l,m\right)$ in all other cases. Examples of
distorted folds are displayed in Fig.~\ref{fig:nc-sol-distorted}
for $L=40\pi$. Solutions which do not satisfy $q=1$, e.g., $\phi_{n=39,m}$
or $\phi_{n=41,m}$ in a $L=40\pi$ domain, obey a similar rule for
the local phase as that obtained by swapping the integer $l$ for
$n$, i.e., replacing $l$ in $\varphi_{f}$ by $n$ to obtain $\phi_{n\neq l,m}$.
Further analysis for this general case is required to understand this
feature.

\subsection{Discussion\label{sub:Discussion}}

\subsubsection{Energy gap}

\begin{figure}
\begin{centering}
\includegraphics{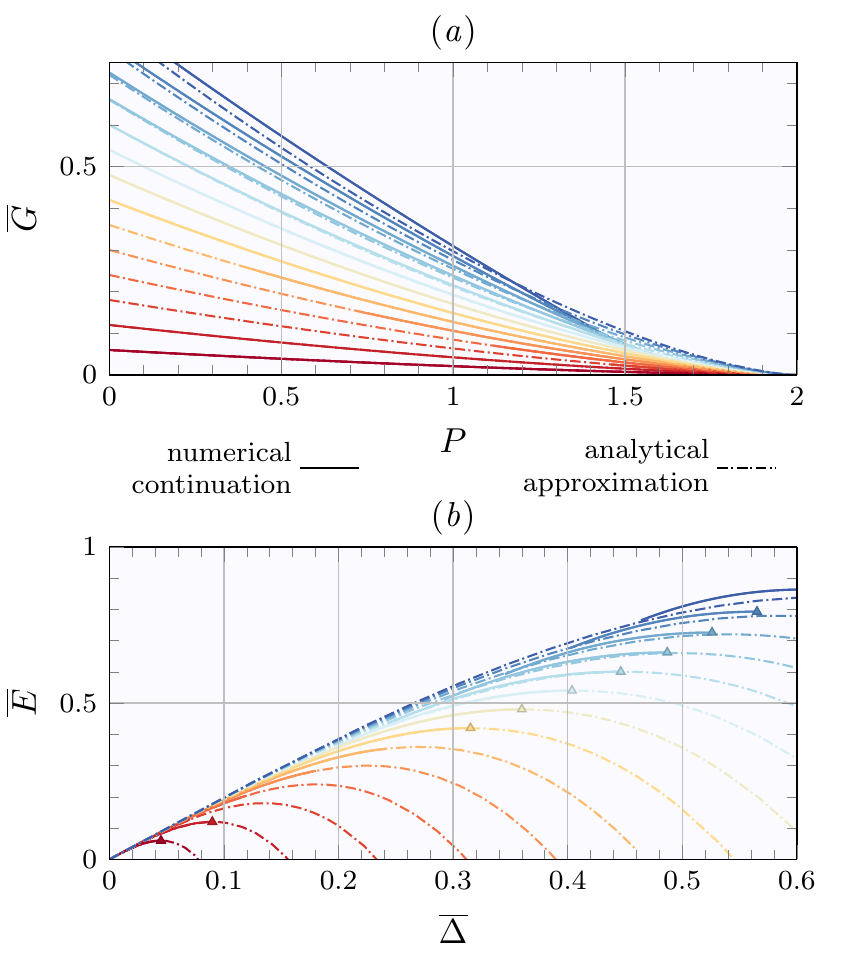} 
\par\end{centering}
\caption{(Color online) Multifold branches for a sheet of length $L=40\pi$
over a fluid substrate in (\textit{a}) the $\left(P,\overline{G}\right)$
plane and (\textit{b}) the $\left(\overline{\Delta},\overline{E}\right)$
plane. Solid lines represent the numerical continuation results while
the dash-dotted lines indicate the leading order analytical approximation.
The numerical continuation was stopped at $P=0$ {[}indicated by $\blacktriangle$
in (\textit{b}){]} in all cases, although the analytical approach
extends beyond this point. The analytical approximation starts to
depart from the numerical results only when the number of folds is
very high, $m>10$. \label{fig:energy}}
\end{figure}

The energy of the dnoidal and cnoidal branches for $l\in\mathbb{Z}^{+}$
and $n=l$ far from the bifurcation can be found by expanding the
expressions for $P$, $\overline{G}$, $\overline{\Delta}$ and $\overline{E}$
obtained in Sec.~\ref{sub:Branches-in-parameter} around $k=1$.
If we write $k=1-\xi$, $\xi\ll1$, we obtain 
\begin{eqnarray}
g_{m}\equiv\frac{1}{m}G_{m}^{\pm}\left(P\right) & = & \frac{8}{3}\left(2-P\right)^{3/2}\left(1\mp\frac{3}{2}\xi\right),\label{eq:G_approx}\\
e_{m}\equiv\frac{1}{m}E_{m}^{\pm}\left(\Delta\right) & = & 2\delta_{m}-\frac{1}{48}\delta_{m}^{3}\left(1\mp3\xi\right),\label{eq:E_approx}
\end{eqnarray}
where the $-$ ($+$) corresponds to the dnoidal (cnoidal) branch
and $\delta_{m}\equiv\Delta/m$ is a single-fold average compression.
Notice that the results are no longer expressed in terms of mean values.
At leading order both branches have exactly the same energy. Equation~(\ref{eq:G_approx})
shows that, for fixed $P$ (dead loading), the $G$ spectrum is a
set of equispaced free energies, where the quantity $g_{m}\approx\frac{8}{3}\left(2-P\right)^{3/2}$
can be identified with the free energy of a single fold. On the other
hand, for fixed compression $\Delta$ (rigid loading), the energy
of a single fold is $e_{m}\approx2\delta-\delta^{3}/48$. The energies
display maxima at $\delta_{\max}=2^{5/2}$, with $e_{\max}=2^{9/2}/3$.
Numerical continuation shows that this point is related to self-contact
of the sheet. A comparison between the numerical results and the analytical
approach at leading order is shown in Fig.~\ref{fig:energy} for
$q=1$. The overlap between the numerical continuation results and
the analytical approximation is remarkable. 

\begin{figure}
\begin{centering}
\includegraphics{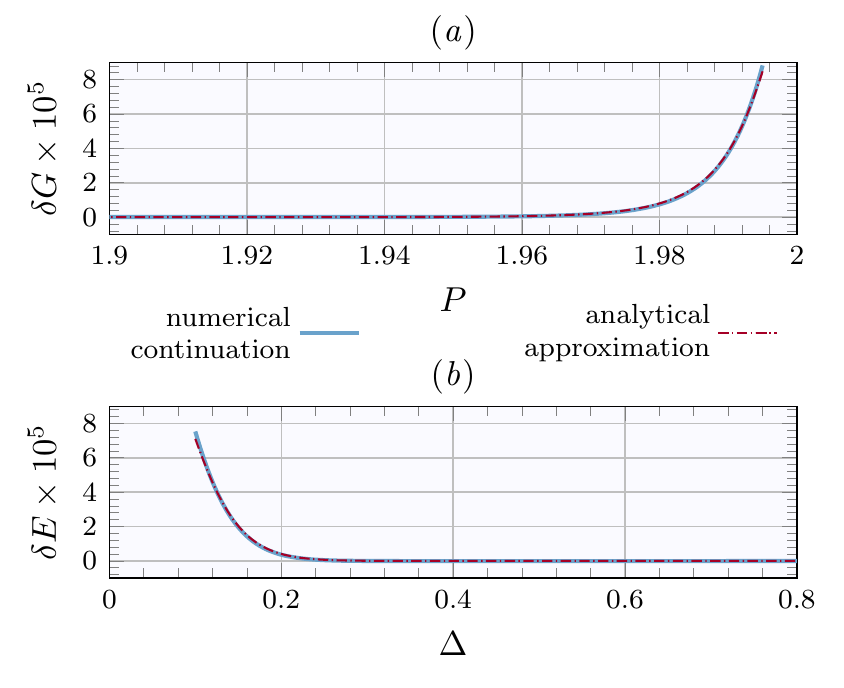} 
\par\end{centering}
\caption{(Color online) Gap between the dnoidal and cnoidal single fold solutions
for $L=40\pi$ in (\textit{a}) the $\left(P,G\right)$ plane and (\textit{b})
the $\left(\Delta,E\right)$ plane. The analytical approximation (dash-dotted
line) once again matches the numerical continuation results (solid
line). \label{fig:difference}}
\end{figure}

Figure~\ref{fig:difference} shows the difference between the $m=1$
dnoidal and cnoidal branches obtained from numerical continuation
and compares the result with a theoretical prediction obtained by
inverting the series expansions for either $P$ or $\Delta$ to obtain
$\xi$ in terms $P$ or $\Delta$. The final result for the energy
gap between the dnoidal and cnoidal branches is 
\begin{eqnarray}
\delta g_{m} & \equiv & g_{m}^{+}-g_{m}^{-}=-64\left(2-P\right)^{\frac{3}{2}}\mathrm{e}^{-\frac{1}{2}l_{m}\left(2-P\right)^{\frac{1}{2}}},\label{eq:G_diff}\\
\delta e_{m} & \equiv & e_{m}^{+}-e_{m}^{-}=-\delta_{m}^{3}\mathrm{e}^{-\frac{1}{8}l_{m}\delta_{m}},\label{eq:E_diff}
\end{eqnarray}
where $l_{m}\equiv L/m$ can be identified as the spacing between
the folds. The predicted exponential decay in both $\left(P,G\right)$
and $\left(\Delta,E\right)$ planes matches the numerical results
almost perfectly. The exponentially small residual difference is a
consequence of the finite sheet size and can be understood as the
interaction between the exponential tails of localized solutions \citep{KZ1}.

The general case $q\neq1$ far from the bifurcation is more complicated
and is not well described by the weak modulation approach. Numerical
continuation shows that, at leading order, the energies for $q\neq1$
are in fact exactly the same as those for the $q=1$ branches: the
solutions appear to shift their $q$ towards $q=1$ as $k\rightarrow1$,
a feature that is beyond the scope of the approach in Sec.~\ref{sub:Approximation}.

\subsubsection{Stability analysis\label{subsec:Stability-analysis}}

The analysis carried out in the previous sections provides a framework
for understanding how single fold and multifold solutions bifurcate
from periodic states. However, this type of analysis does not address
the stability of the solutions, although we expect solutions with
lowest energy such as the single fold solution $\phi_{40,1}$ for
$L=40\pi$ to be stable. For example, it may happen that several solutions
are stable simultaneously. Thus a stability analysis provides a link
between the theoretical predictions and observations in experiments.

To explore the stability properties of the solutions identified above,
we carried out a numerical study based on a variational approach.
Since the equations of a fluid-supported elastic sheet are static
(time-independent), we consider a solution as stable if and only if
it corresponds to a local minimum of energy in the space of functions
satisfying all the constraints, as well as the boundary and symmetry
conditions imposed by the problem formulation. The stability of a
solution is thus guaranteed if the energy functional is positive definite
in its vicinity, i.e., if and only if $E\left[\phi_{0}+\delta\phi\right]>E\left[\phi_{0}\right],\;\forall\,\delta\phi\,\in\,\mathbb{F}_{\phi}$,
where $E$ is defined as in Eq.~\eqref{eq:total_energy}, $\phi_{0}$
is the solution under study, $\delta\phi\neq0$ is the variation of
$\phi_{0}$ and $\mathbb{F}_{\phi}$ is the space of (continuous)
functions such that $\phi_{0}+\delta\phi$ satisfies the prescribed
constraints (e.g. $\Delta\left[\delta\phi\right]=0$ under rigid loading),
and the boundary and symmetry conditions (hinged, symmetric or antisymmetric).

We define the inner product $\left(f,g\right)\equiv\int_{-L/2}^{L/2}f\cdot g\,\mathrm{d}s=\int_{-L/2}^{L/2}f^{\dagger}g\,\mathrm{d}s$.
It can be shown (see Appendix~\ref{sub:Appendix-B}) that the functionals
$E,G$ evaluated at $y=y_{0}+\delta y$ can be written as 
\begin{align*}
E\left[y\right]= & E_{0}+\left(\frac{\delta E}{\delta y},\delta y\right)+\left(\delta y,\frac{\delta^{2}E}{\delta y}\delta y\right)+\mathcal{O}\left(\delta y^{3}\right),\\
G\left[y\right]= & G_{0}+\left(\frac{\delta G}{\delta y},\delta y\right)+\left(\delta y,\frac{\delta^{2}G}{\delta y}\delta y\right)+\mathcal{O}\left(\delta y^{3}\right),
\end{align*}
where $E_{0}=E\left[\phi_{0}\right]$, $G_{0}=G\left[\phi_{0}\right]$
and all the functional derivatives are evaluated at $\phi_{0}$ (equivalently,
$y=y_{0}$). The variation $\delta y$ is related to the corresponding
variation $\delta\phi$ by the geometrical relation $\partial_{s}\left(y+\delta y\right)=\sin\left(\phi+\delta\phi\right)$
(see Appendix~\ref{sub:Appendix-B} for further details). Similar
relations can be written for $P$ and $\Delta$. By construction,
the terms in $E$ or $G$ linear in $\delta y$ vanish when $y=y_{0}$
is a solution of the original variational problem. Hence, under dead
loading ($P$ fixed) the solution is stable provided 
\[
\left(\delta y,\frac{\delta^{2}G}{\delta y^{2}}\delta y\right)=\left(\delta y,\left[\frac{\delta^{2}E}{\delta y^{2}}-P_{0}\frac{\delta^{2}\Delta}{\delta y^{2}}\right]\delta y\right)>0.
\]
Here $P_{0}=P[y_{0}]$. In the rigid loading case, $P$ is no longer
fixed and the stability condition becomes 
\[
\left(\delta y,\left[\frac{\delta^{2}E}{\delta y^{2}}-P_{0}\frac{\delta^{2}\Delta}{\delta y^{2}}\right]\delta y\right)-\delta P\left(\frac{\delta\Delta}{\delta y},\delta y\right)>0
\]
with an extra constraint imposed by the fixed compression: 
\[
\left(\frac{\delta\Delta}{\delta y},\delta y\right)=0.
\]
Since the operator $\delta^{2}G/\delta y^{2}$ in the square brackets
is a real symmetric operator, the inner products behave as a quadratic
form in function space. Thus, the spectral decomposition of the operator
yields a set of eigenvalues and the corresponding eigenfunctions containing
information about the stability of the solution.

In practice, we first recovered $\phi_{0}$ and the corresponding
$y_{0}$ from the AUTO$^{\text{©}}$ calculations and resampled them
into an $N$ vector. The derivative operators from Eqs.~\eqref{eq:dDelta}-\eqref{eq:dEnergy}
were likewise discretized into $N\times N$ arrays using Fourier derivative
matrices. Using these results, we built a discrete $N\times N$ matrix
of the free energy functional $\delta^{2}G/\delta y^{2}$, $\mathbf{G}_{N\times N}$.
The stability of solutions in the dead loading case simply depends
on the spectral decomposition of $\mathbf{G}_{N\times N}$, that is,
on the eigenvalues of the standard eigenvalue problem 
\begin{equation}
\mathbf{G}\left[\delta y\right]=\lambda\left[\delta y\right].\label{eq:stabilityDL}
\end{equation}

The rigid loading case requires further work because of the extra
constraint. In this case we discretized the compression condition
by introducing a matrix $\Delta_{N\times1}'$ as the discrete version
of $\delta\Delta/\delta y$, and the ${\left(N+1\right)\times\left(N+1\right)}$
bordered Hessian matrix $\mathbf{H}$ given by 
\[
\mathbf{H}\equiv\left[\begin{array}{rr}
0 & -\mathbf{\Delta}_{1\times N}'\\
-\mathbf{\Delta}_{N\times1}' & \mathbf{G}_{N\times N}
\end{array}\right].
\]
As shown in \citep{chiang2013fundamental}, the stability problem
can be studied via the generalized eigenvalue problem 
\begin{equation}
\mathbf{H}\left[\begin{array}{c}
\delta P\\
\delta y
\end{array}\right]=\lambda\left[\begin{array}{ll}
0 & \bm{0}_{1\times N}\\
\bm{0}_{N\times1} & \bm{I}_{N\times N}
\end{array}\right]\left[\begin{array}{c}
\delta P\\
\delta y
\end{array}\right].\label{eq:stabilityRL}
\end{equation}
Thus the stability of a solution under dead or rigid loading can be
tested by checking the signs of the eigenvalues of the standard (Eq.~\ref{eq:stabilityDL})
or generalized eigenvalue (Eq.~\ref{eq:stabilityRL}) problems, respectively.
If all the eigenvalues are positive, linear stability is guaranteed.

\begin{figure}[b]
\begin{centering}
\includegraphics{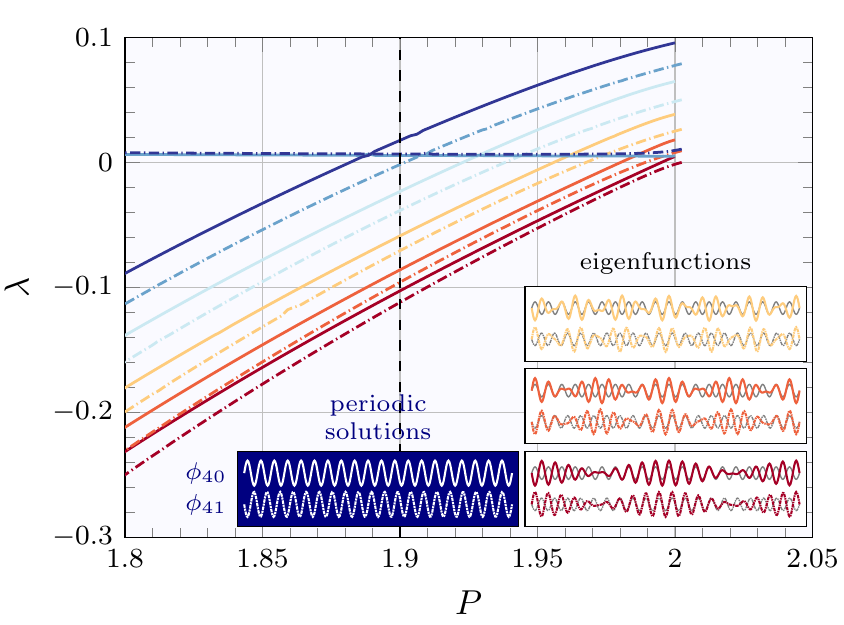} 
\par\end{centering}
\caption{(Color online) Eigenvalues $\lambda$ as a function of the load $P$
for two periodic solutions: $\phi_{40}$ (solid lines) and $\phi_{41}$
(dash-dotted lines). The left inset shows the periodic solutions evaluated
at $P=1.9$ (vertical dashed line in main panel). The right insets
show the eigenfunctions evaluated at the same value of $P$ using
the same color scheme as in the main panel. The $\phi_{40}$ eigenfunctions
are displayed at the top (thick lines) and the $\phi_{41}$ eigenfunctions
at the bottom (thin lines). The corresponding periodic solutions are
plotted in black in the background for reference. \label{fig:eigen_periodic}}
\end{figure}

Within this framework, we now proceed to describe the stability properties
of the solutions for a $L=40\pi$ sheet. Under dead loading, all the
solutions described so far are unstable to amplitude modes, which
is as expected as $\partial G/\partial P<0$ for all the branches
studied. The rigid loading case is much more interesting, however.
In the following section, we describe the stability properties of
the periodic solutions $\phi_{40}$ and $\phi_{41}$, the symmetric
and asymmetric single fold solutions $\phi_{40,1}$ and $\phi_{41,1}$,
and the two-fold solutions $\phi_{39,2}$, $\phi_{40,2}$, $\phi_{41,2}$
and $\phi_{42,2}$. For the sake of completeness, we also examined
the stability of solutions displaying half-folds at the boundaries,
which we denote by the superscript $b$. For instance, the solution
$\phi_{40,1}^{\left(b\right)}$ has half-folds at $s=\pm L/2$, i.e.,
$\frac{1}{2}$+$\frac{1}{2}$=1 complete fold. \\

\begin{figure}[t]
\begin{centering}
\includegraphics{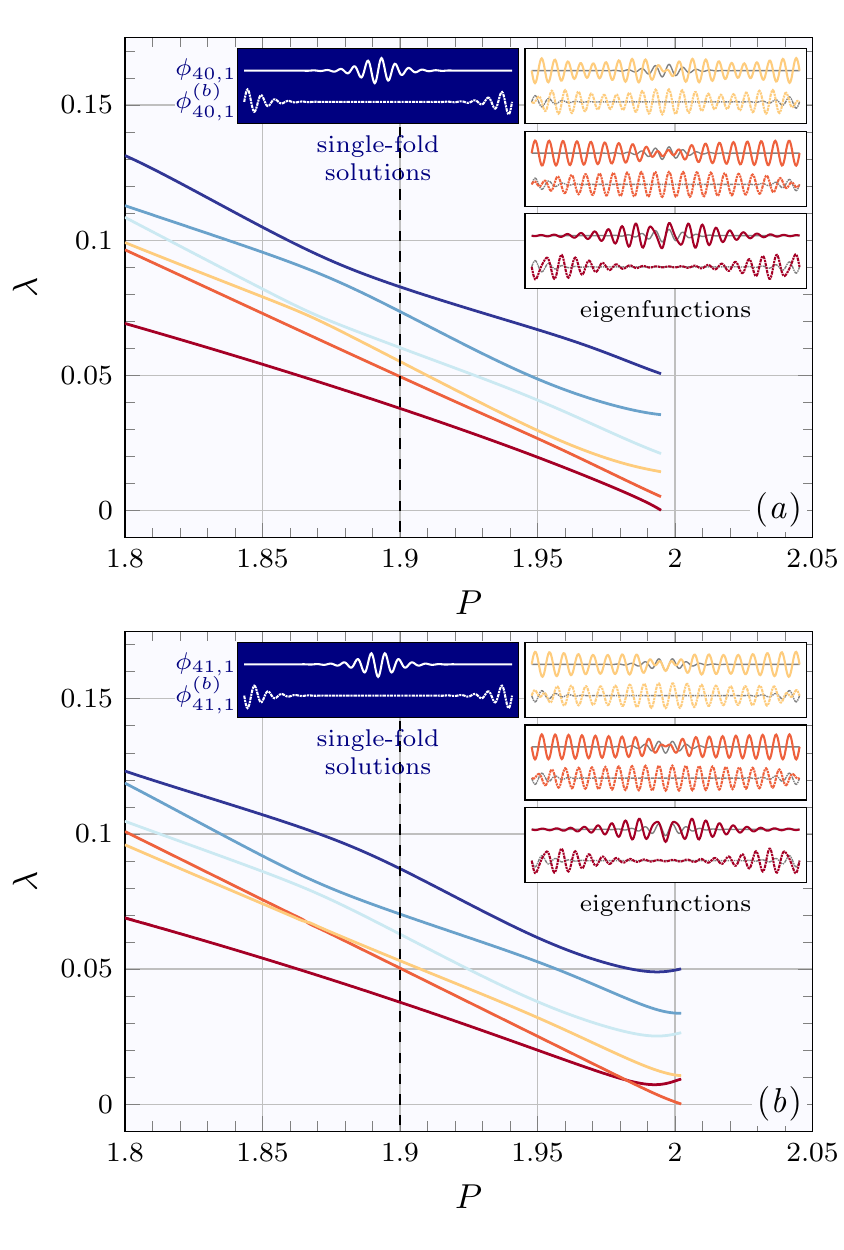} 
\par\end{centering}
\caption{(Color online) Eigenvalues $\lambda$ as a function of the load $P$
for four single fold solutions. (\emph{a}) Symmetric fold: $\phi_{40,1}$
(solid lines) and $\phi_{40,1}^{\left(b\right)}$ (dash-dotted lines).
(\emph{b}) Antisymmetric fold: $\phi_{41,1}$ (solid lines) and $\phi_{41,1}^{\left(b\right)}$
(dash-dotted lines). As in Fig.~\ref{fig:eigen_periodic}, the left
insets show the solutions evaluated at $P=1.9$ and those on the right
show the eigenfunctions. The centered fold eigenfunctions are displayed
at the top (solid lines) and the boundary half-fold eigenfunctions
at the bottom (dashed lines). \label{fig:eigen_single-fold}}
\end{figure}

\subsubsection{Stability of periodic solutions\label{subsec:Stability-of-periodic}}

Figure~\ref{fig:eigen_periodic} shows the six lowest eigenvalues
$\lambda$ as a function of the load $P$ for the periodic solutions
$\phi_{40}$ (solid lines) and $\phi_{41}$ (dashed lines). The corresponding
eigenfunctions evaluated at $P=1.9$ are shown in the right insets
following the same color scheme as used for the eigenvalues. The figure
indicates that the periodic solution $\phi_{40}$ becomes unstable
at $P=1.995$, when its lowest eigenvalue becomes negative. The corresponding
eigenfunction represents a buckling mode that leads to either a centered
single fold state or a pair of half-folds on the boundaries. We will
refer to these eigenfunctions as fold eigenmodes. At $P\approx1.980,$
a second eigenvalue becomes negative, which buckles the solution into
a pair of folds (or a $\frac{1}{2}$+1+$\frac{1}{2}$-fold state);
at $P\approx1.956$, a third instability sets in, buckling the solution
into three folds or a $\frac{1}{2}$+1+1+$\frac{1}{2}$-fold state,
etc. The points where these solutions cross $\lambda=0$ are consistent
with the branch points identified using AUTO$^{\text{©}}$ as leading
to solutions that subsequently evolve into fold solutions. Similar
behavior is observed for the periodic solution $\phi_{41}$, with
fold eigenvalues crossing $\lambda=0$ sequentially as the number
of folds increases. Besides the fold eigenvalues all of which decrease
monotonically as $P$ decreases, a second family of eigenvalues is
also present. However, these never cross $\lambda=0$ even though
they decrease monotonically as $P$ decreases. The associated eigenfunctions
do not buckle the solution into localized folds but instead introduce
spatial modulation on top of the periodic pattern. We conclude that
the periodic solutions are stable with respect to spatial modulation,
at least when the wavelength of the periodic state is close to the
natural wavelength (no Eckhaus instability).

Figure \ref{fig:eigen_types}(\emph{a}) shows the construction of
a $n=40$ pseudo-solution arising from the second unstable fold mode.
The pseudo-solution is obtained by adding a multiple of the eigenfunction
(here the multiple is $\pm1$) to the periodic state $\phi_{40}$.
Panel (\textit{a}) shows that when the multiple is $+1$ the instability
suppresses the oscillations in the center and near the boundaries,
leaving a $1+1$ state. When the multiple is $-1$ the result is a
state resembling a $\frac{1}{2}$+1+$\frac{1}{2}$ state.

\subsubsection{Stability of single fold solutions\label{subsec:Stability-of-single}}

The corresponding diagram for single fold solutions is shown in Fig.~\ref{fig:eigen_single-fold}.
The four single fold solutions have positive eigenvalues and hence
are all stable. The first two solutions $\phi_{40,1}$ and $\phi_{40,1}^{\left(b\right)}$,
have identical eigenvalues, a consequence of the fact that $\phi_{40,1}^{\left(b\right)}$
can be obtained through a half-domain translation of $\phi_{40,1}$.
This translation respects the boundary conditions at $s=\pm L/2$.
The lowest eigenvalue of $\phi_{40,1}$ is inherited from the periodic
branch $\phi_{40}$ and represents the amplitude mode of $\phi_{40}$
{[}Fig.~\ref{fig:eigen_types}(\textit{b}){]}. 
The second lowest eigenvalue is of particular interest, because its
eigenmode is related to the splitting of the central fold into two
{[}Fig.~\ref{fig:eigen_types}(\textit{c}){]}. These eigenmodes,
which will be referred to as splitting eigenmodes, represent modulation
of a carrier wave that shifts continuously with respect to the base
state, with a $\pi$-phase shift at the center of the fold. With increasing
amplitude this mode leads to growing separation of the central fold
into two adjacent folds, in contrast to the usual fold eigenmodes
corresponding to a $0$ or $\pi$ shift. The third eigenvalue is linked
to a fold eigenmode which buckles the solution into a $\frac{1}{2}$+$1$+$\frac{1}{2}$
array. Larger eigenvalues show similar behavior involving larger numbers
of folds. However, none of these modes is unstable, and so no finite
amplitude solutions of this type are present.

The stability behavior the $\phi_{41,1}$ state is similar to that
of $\phi_{40,1}$ although there are some important differences. First,
the solution $\phi_{41,1}^{\left(b\right)}$ no longer corresponds
to a simple half-domain shift of $\phi_{41,1}$: to satisfy symmetry
conditions, a $\frac{\pi}{2}$ phase shift of the carrier wave plus
an inversion of the solution in one half of the domain is also required.
As a result the eigenvalues of $\phi_{41,1}$ and $\phi_{41,1}^{\left(b\right)}$
are no longer identical. The slight difference between them ($\lesssim10^{-4}$)
is due to the presence of a defect in the center of the domain in
the latter case. The lowest eigenvalue mode is again related to the
periodic solution, $\phi_{41,0}$. The next lowest eigenvalue mode
is now a $1$+$1$ eigenmode while the third is a splitting eigenmode.
Notice also that the second mode has the lowest eigenvalue for $P\lessapprox1.982$
but this eigenvalue, like the others, remains positive and no instability
is triggered.

\begin{figure*}[t]
\begin{centering}
\includegraphics{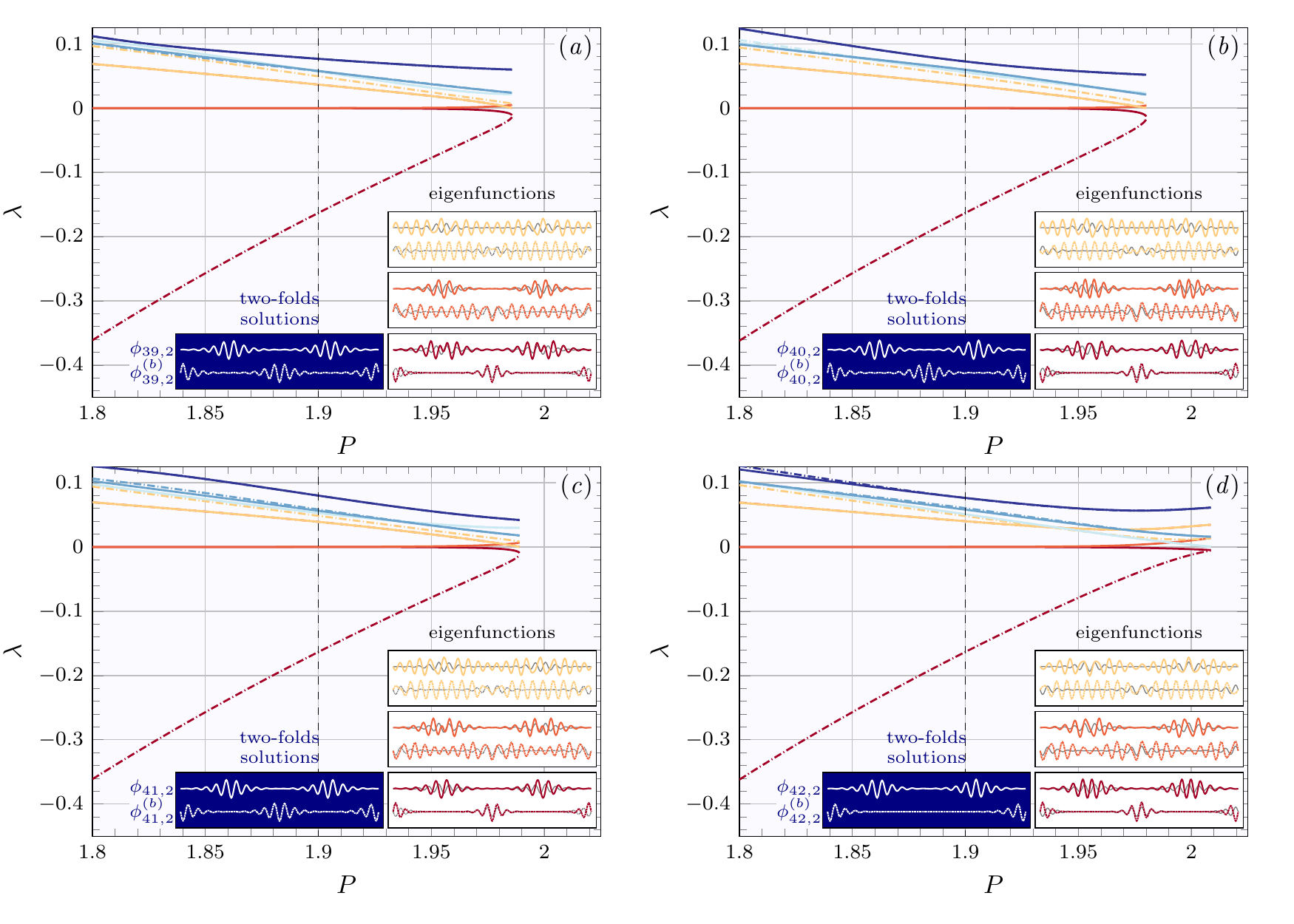} 
\par\end{centering}
\caption{(Color online) Eigenvalues $\lambda$ as a function of the load $P$
for the eight two-fold solutions. (\emph{a}) $\phi_{39,2}$ (solid
lines) and $\phi_{39,2}^{\left(b\right)}$ (dash-dotted lines). (\emph{b})
$\phi_{40,2}$ (solid lines) and $\phi_{40,2}^{\left(b\right)}$ (dash-dotted
lines). (\emph{c}) $\phi_{41,2}$ (solid lines) and $\phi_{41,2}^{\left(b\right)}$
(dash-dotted lines). (\emph{d}) $\phi_{42,2}$ (solid lines) and $\phi_{42,2}^{\left(b\right)}$
(dash dotted lines). As in Fig.~\ref{fig:eigen_periodic}, the left
insets show the solutions evaluated at $P=1.9$ and those on the right
show the eigenfunctions. The centered fold eigenfunctions are displayed
at the top (solid lines) and boundary half-fold eigenfunctions at
the bottom (dashed lines). \label{fig:eigen_double-fold}}
\end{figure*}

\subsubsection{Stability of two-fold solutions\label{subsec:Stability-of-two-folds}}

\begin{figure}[t]
\begin{centering}
\includegraphics{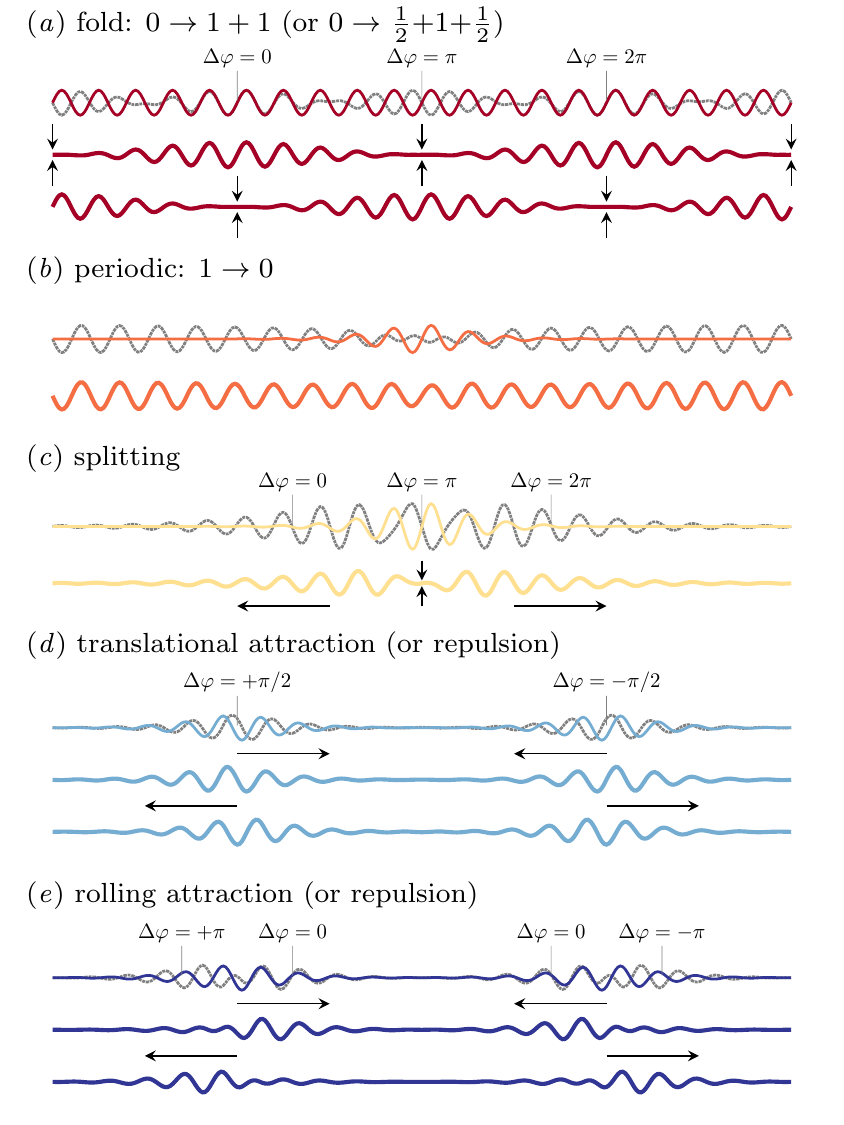} 
\par\end{centering}
\caption{(Color online) Examples of base states (top solid lines) and the corresponding
eigenmodes (top dash-dotted lines) and pseudo-solutions obtained by
their superposition (bottom lines). (\textit{a}) Periodic solution
(red) and the fold eigenmode (black) superposed in two ways. (\textit{b})
Single fold solution (orange) and the periodic eigenmode (gray). (\textit{c})
Single fold solution (yellow) and the splitting eigenmode (gray).
(\textit{d}) Two-fold solution (gray) and the translational attraction/repulsion
eigenmode (blue). (\textit{e}) Two-fold solution (blue) and the rolling
attraction/repulsion eigenmode (gray).\label{fig:eigen_types}}
\end{figure}

We also studied the stability properties of the two-fold solutions.
The eigenvalues of the $\phi_{39,2}$, $\phi_{40,2}$, $\phi_{41,2}$
and $\phi_{42,2}$ solutions are shown in Fig.~\ref{fig:eigen_double-fold}
(solutions with folds on the boundaries are also shown). All the two-fold
solutions are unstable displaying a single negative eigenvalue. At
the bifurcation point to a two-fold solution (e.g. at $P\approx1.98$
for $\phi_{39,2}$) the amplitude eigenvalue inherited from the periodic
state vanishes but becomes positive as $P$ decreases just as in the
single fold case. However, this eigenvalue is passed by that of another
mode in the vicinity of the bifurcation point (e.g. at $P\lessapprox1.975$
for $\phi_{39,2}$). The corresponding eigenmode is non-periodic and
remarkably interesting. For the $\phi_{41,2}$ and $\phi_{42,2}$
solutions, these two lowest eigenvalue modes display the same modulation
as the periodic solution but with a $\pm\frac{\pi}{2}$ shift in the
carrier wave. Because of the localization of the folds, this feature
is a reminiscent of a Goldstone translation mode. The eigenmode acts
on the solutions by translating the left fold to the right ($+\frac{\pi}{2}$
shift) and the right fold to the left ($-\frac{\pi}{2}$ shift), or
vice versa, without changing their shape. Consequently we refer to
this mode as an translational attraction (or repulsion) eigenmode.
The other lowest eigenvalue mode is slightly different: on a one side
of a fold the mode displays a $\pi$ shift while on the other side
there is no phase shift. These eigenmodes also generate attraction
(or repulsion) between folds but also change their shape. We refer
to them as rolling attraction (or repulsion) eigenmodes. While in
the translation mode both the carrier wave and the envelope translate,
in the rolling case only the envelope shifts, resulting in a local
deformation of the folds. The situation for $\phi_{39,2}$ and $\phi_{40,2}$
is similar with the difference that the translation and rolling eigenmodes
change their order. Figures~\ref{fig:eigen_types}(\emph{d}) and
(\emph{e}) show pseudo-solutions that indicate the tendencies represented
by these modes. We conclude that the two-fold states are in all cases
unstable and that the instability takes the form of a gradual approach
of the two folds, ultimately resulting in a stable single fold state.
The manifestation of the instability may be very slow, however, as
it depends on the energy difference between these two states. This
conclusion is supported by the fact that the unstable eigenvalue of
the two-fold state approaches zero from below, asymptotically exponentially,
as $P$ decreases. This comes as no surprise since the two folds can
only interact via the overlapping tails of their profiles and these
are exponentially small when the folds are sufficiently far apart,
a fact that suggests that as the solutions increasingly localize (i.e.,
$P$ decreases), the time required for the system to reach the single
fold state becomes exponentially long. This decay may arise through
attraction via either the translation or rolling eigenmodes. In contrast,
two-fold solutions with folds on the boundary decay into a single
fold state much more rapidly as the folds at the boundary decay leaving
only the fold at the center.

\section{Conclusions}

In this article, we reported on the existence of multifold solutions
in a thin floating elastic sheet of finite length with hinged boundary
conditions. Starting from known large amplitude periodic solutions
\cite{Oshri:2015bi}, we used numerical continuation to uncover a
series of solution branches that bifurcate from the periodic (wrinkle)
branch as a function of the load $P$ or the associated compression
$\Delta$. Each branch consists of an array of localized (fold) solutions.
Symmetric and antisymmetric single fold solutions, which are well
known for the infinite sheet case \cite{Diamant:2013je}, are found
to bifurcate from branches of distinct periodic states and inherit
their symmetry properties. The multifold states with $m=2,3,\dots$
folds in the domain bifurcate from the periodic states in subsequent
bifurcations and are, we believe, new.

Based on our numerical continuation results, we performed a weakly
nonlinear analysis of the multifold solutions. These states form as
a result of a slow spatial modulation of the periodic wrinkle solutions.
The spatial modulation satisfies an amplitude equation whose solutions
can be written down in terms of Jacobi elliptic functions. The results
determine the branches of multifold solutions in the parameter space
and match the numerical continuation results remarkably well. In particular,
we determined the energy of multifold states and showed that the energies
of like-number multifold states are almost equal, with energy gaps
that become exponentially small as the sheet length increases (see
Eq.~(\ref{eq:E_approx})).

Finally, we also studied the stability of both the periodic and multifold
solutions within a variational framework. We showed numerically that
the symmetric and antisymmetric single fold solutions are both stable,
but that only one of them bifurcates from a stable periodic branch
and so inherits its stability. We also showed that all the two-fold
solutions are unstable, but with a common instability mode characterized
by the mutual attraction/repulsion of folds (see Sec.~\ref{fig:eigen_double-fold}).
However, the relevant eigenvalue comes very close to zero as the localization
increases, implying that the transition to a single fold state requires
essentially exponentially long times. This result is in agreement
with the numerical simulations in \cite{Marple:2015kq} (with clamped
boundary conditions) in which a three-fold state is observed to decay
into a single fold state but did so so slowly that it was thought
to be stable.

These results have been obtained for the special case of hinged boundary
conditions. However, some numerical experiments employ clamped boundary
conditions, see e.g. \cite{Rivetti:2014dh}, and in this case the
route to localization appears to be different. Thus an extension of
the type of analysis performed here to the case of clamped boundary
conditions $\phi=\phi''=0$ at $s=\pm L/2$ would also be valuable.

Our results can be used to provide a rule of thumb for experiments
employing hinged boundary conditions where localization of wrinkles
into folds is crucial. Multifold solutions may emerge naturally in
certain geometries (for instance, in a floating elastic annulus or
an indented circular sheet \cite{Paulsen:2016hm,Paulsen:2017dq}).
The prediction of the energy gap between multifold and single fold
states obtained here can be used to determine which states can emerge
deterministically or stochastically in the presence of thermal noise.
This is of particular importance in microscopic-scale experiments
\cite{Gopal:2006gk,Leahy:2010he} and in experiments in which the
floating raft is granular \cite{JambonPuillet:2017fa}. Although in
most experiments folds are allowed to form spontaneously, our calculations
also provide a basis for predicting the outcome of experiments with
manually induced folds, which may be of particular interest in technological
applications. 
\begin{acknowledgments}
This work was partially funded by the Berkeley-Chile Fund for collaborative
research and CONICYT-USA PII20150011. L.G. wishes to acknowledge support
by Conicyt/Becas Chile de Postdoctorado 74150032 and Conicyt PAI/IAC
79160140. We are grateful to Enrique Cerda, Marcel Clerc and Punit
Gandhi for fruitful discussions. 
\end{acknowledgments}

\appendix

\section{Jacobi elliptic integrals\label{sub:Appendix-A}}

The quantities related to $A$ in Sec.~\ref{sub:Branches-in-parameter}
can be written in terms of fundamental Jacobi elliptic integrals,
\begin{eqnarray*}
\left(\overline{A\,_{\dn}^{2}},\overline{A\,_{\cn}^{2}}\right) & = & 2^{4}\kappa^{2}\left(D_{2},C_{2}\right),\\
\left(\overline{A_{\dn}^{'2}},\overline{A_{\cn}^{'2}}\right) & = & 2^{4}\kappa^{4}(D{}_{2}^{'},C{}_{2}^{'}),\\
\left(\overline{A\,{}_{\dn}^{4}},\overline{A\,{}_{\cn}^{4}}\right) & = & 2^{8}\kappa^{4}\left(D_{4},C{}_{4}\right),
\end{eqnarray*}
where the coefficients $D_{2},C_{2},D{}_{2}^{'},C{}_{2}^{'},D_{4},C_{4}$
are given, respectively, by the following mean values of the Jacobi
elliptic functions: 
\[
\begin{array}{ccccl}
D_{2} & \equiv & \overline{\dn_{k}^{2}} & = & 1-I(k)\\
D_{2}^{'} & \equiv & \overline{\dn_{k}^{'2}} & = & \frac{1}{3}\left[k^{2}-\left(2-k^{2}\right)I(k)\right]\\
D_{4} & \equiv & \overline{\dn_{k}^{4}} & = & \frac{1}{3}\left[\left(3-k^{2}\right)-2\left(2-k^{2}\right)I(k)\right]\\
C_{2} & \equiv & \overline{k^{2}\cn_{k}^{2}} & = & k^{2}-I(k)\\
C_{2}^{'} & \equiv & \overline{k^{2}\cn_{k}^{'2}} & = & \frac{1}{3}\left[k^{2}-\left(2k^{2}-1\right)I(k)\right]\\
C_{4} & \equiv & \overline{k^{4}\cn_{k}^{4}} & = & \frac{1}{3}\left[k^{2}\left(3k^{2}-1\right)-2\left(2k^{2}-1\right)I(k)\right].
\end{array}
\]

\section{Expansion of functionals\label{sub:Appendix-B}}

Consider a small perturbation of a solution $y_{0}$ satisfying the
variational problem, $y=y_{0}+\delta y$, where $\delta y$ is assumed
to be of order $\varepsilon$, i.e. $\delta y=\varepsilon y_{1}\neq0$
and $y_{1}\sim\mathcal{O}\left(1\right)$: 
\begin{eqnarray*}
y & = & y_{0}+\varepsilon y_{1}.
\end{eqnarray*}
The function $\phi\left(s\right)$ can be expanded in terms of $\varepsilon$,
\[
\phi=\phi_{0}+\varepsilon\phi_{1}+\varepsilon^{2}\phi_{2}+\ldots
\]
and likewise for the load $P$, 
\begin{eqnarray*}
P & = & P_{0}+\varepsilon P_{1}+\ldots
\end{eqnarray*}
and the compression and energy functionals, 
\begin{eqnarray*}
\Delta & = & \Delta_{0}+\varepsilon\Delta_{1}+\varepsilon^{2}\Delta_{2}+\ldots\\
E & = & E_{0}+\varepsilon E_{1}+\varepsilon^{2}E_{2}+\ldots\\
G & = & G_{0}+\varepsilon G_{1}+\varepsilon^{2}G_{2}+\ldots
\end{eqnarray*}
Each term in these expansions can be obtained from the definitions
of these relevant quantities (Sec.~\ref{subsec:Governing-equations}).
For instance, the function $\phi$ is related to $y$ through $\dot{y}=\sin\phi$,
and it can be shown that the first and second order terms in the expansion
of $\phi$ are, respectively, 
\begin{eqnarray*}
\phi_{1} & = & \sec\phi_{0}\,\partial_{s}y_{1},\\
\phi_{2} & = & \frac{1}{2}\tan\phi_{0}\sec^{2}\phi_{0}\left(\partial_{s}y_{1}\right)^{2}.
\end{eqnarray*}
Analogously, the functionals $\Delta$ and $E$ can be written as
\begin{align*}
\Delta & =\Delta_{0}+\varepsilon\left(\frac{\delta\Delta}{\delta y},y_{1}\right)+\varepsilon^{2}\left(y_{1},\frac{\delta^{2}\Delta}{\delta y}y_{1}\right)\\
E & =E_{0}+\varepsilon\left(\frac{\delta E}{\delta y},y_{1}\right)+\varepsilon^{2}\left(y_{1},\frac{\delta^{2}E}{\delta y}y_{1}\right),
\end{align*}
where $\Delta_{0}$ and $E_{0}$ and their functional derivatives
are evaluated at the solution $y_{0}$. For the sake of simplicity,
we only consider here the terms that are relevant for the stability
analysis. For the displacement, we obtain: 
\begin{align}
\frac{\delta\Delta}{\delta y} & =\tan\phi_{0}D_{s},\label{eq:dDelta}\\
\frac{\delta^{2}\Delta}{\delta y^{2}} & =D_{s}^{\dagger}\Sigma_{s}^{\Delta}D_{s},\label{eq:d2Delta}
\end{align}
while for the energy, 
\begin{align}
\frac{\delta^{2}E}{\delta y^{2}} & =D_{ss}^{\dagger}\Sigma_{ss}^{E}D_{ss}-D_{s}^{\dagger}\Sigma_{s}^{E}D_{s}+\Sigma_{0}^{E},\label{eq:dEnergy}
\end{align}
where $D_{s}$ and $D_{ss}$ are, respectively, the first and second
order linear differential operators and $D_{s}^{\dagger}$ and $D_{ss}^{\dagger}$
are the corresponding adjoint operators. The scalar operators $\Sigma_{s}^{D}$,
$\Sigma_{ss}^{E}$, $\Sigma_{s}^{E}$ and $\Sigma_{0}^{E}$ are given
by 
\begin{align*}
\Sigma_{s}^{\Delta}\equiv & \frac{1}{2}\sec^{3}\phi_{0},\\
\Sigma_{0}^{E}\equiv & \frac{1}{2}\sec^{2}\phi_{0}\left(y_{0}\dot{\phi}_{0}+\cos\phi_{0}\right),\\
\Sigma_{s}^{E}\equiv & \frac{1}{4}y_{0}^{2}\sec^{3}\phi_{0}+\ldots\\
 & \left[\dot{\phi_{0}}^{2}\left(\frac{1}{2}+\tan^{2}\phi_{0}\right)+\ddot{\phi}_{0}\tan\phi_{0}\right]\sec^{2}\phi_{0},\\
\Sigma_{ss}^{E}\equiv & \frac{1}{2}\sec^{3}\phi_{0}.
\end{align*}
A straightforward calculation shows that the free-energy second-order
term $G_{2}$ is given by 
\[
G_{2}=\left(y_{1},\frac{\delta^{2}G}{\delta y}y_{1}\right)-P_{1}\left(\frac{\delta\Delta}{\delta y},y_{1}\right)
\]
, where the second order functional derivative is simply 
\[
\frac{\delta^{2}G}{\delta y^{2}}=\frac{\delta^{2}E}{\delta y^{2}}-P_{0}\frac{\delta^{2}\Delta}{\delta y^{2}}.
\]
Notice that the functional derivatives have been introduced in symmetrical
form, which simplifies the numerical calculations of Sec.~\ref{subsec:Stability-analysis}.

 \bibliographystyle{apsrev}
\bibliography{localization}

\begin{thebibliography}{35}
\expandafter\ifx\csname natexlab\endcsname\relax\def\natexlab#1{#1}\fi
\expandafter\ifx\csname bibnamefont\endcsname\relax
  \def\bibnamefont#1{#1}\fi
\expandafter\ifx\csname bibfnamefont\endcsname\relax
  \def\bibfnamefont#1{#1}\fi
\expandafter\ifx\csname citenamefont\endcsname\relax
  \def\citenamefont#1{#1}\fi
\expandafter\ifx\csname url\endcsname\relax
  \def\url#1{\texttt{#1}}\fi
\expandafter\ifx\csname urlprefix\endcsname\relax\def\urlprefix{URL }\fi
\providecommand{\bibinfo}[2]{#2}
\providecommand{\eprint}[2][]{\url{#2}}

\bibitem[{\citenamefont{Pocivavsek et~al.}(2008)\citenamefont{Pocivavsek,
  Dellsy, Kern, Johnson, Lin, Lee, and Cerda}}]{Pocivavsek:2008kz}
\bibinfo{author}{\bibfnamefont{L.}~\bibnamefont{Pocivavsek}},
  \bibinfo{author}{\bibfnamefont{R.}~\bibnamefont{Dellsy}},
  \bibinfo{author}{\bibfnamefont{A.}~\bibnamefont{Kern}},
  \bibinfo{author}{\bibfnamefont{S.}~\bibnamefont{Johnson}},
  \bibinfo{author}{\bibfnamefont{B.}~\bibnamefont{Lin}},
  \bibinfo{author}{\bibfnamefont{K.~Y.~C.} \bibnamefont{Lee}},
  \bibnamefont{and} \bibinfo{author}{\bibfnamefont{E.}~\bibnamefont{Cerda}},
  \bibinfo{journal}{Science} \textbf{\bibinfo{volume}{320}},
  \bibinfo{pages}{912} (\bibinfo{year}{2008}).

\bibitem[{\citenamefont{Diamant and Witten}(2011)}]{2011PhRvL.107p4302D}
\bibinfo{author}{\bibfnamefont{H.}~\bibnamefont{Diamant}} \bibnamefont{and}
  \bibinfo{author}{\bibfnamefont{T.~A.} \bibnamefont{Witten}},
  \bibinfo{journal}{Phys. Rev. Lett.} \textbf{\bibinfo{volume}{107}},
  \bibinfo{pages}{164302} (\bibinfo{year}{2011}).

\bibitem[{\citenamefont{Audoly}(2011)}]{Audoly:2011cq}
\bibinfo{author}{\bibfnamefont{B.}~\bibnamefont{Audoly}},
  \bibinfo{journal}{Phys. Rev. E} \textbf{\bibinfo{volume}{84}},
  \bibinfo{pages}{011605} (\bibinfo{year}{2011}).

\bibitem[{\citenamefont{Rivetti and Neukirch}(2014)}]{Rivetti:2014dh}
\bibinfo{author}{\bibfnamefont{M.}~\bibnamefont{Rivetti}} \bibnamefont{and}
  \bibinfo{author}{\bibfnamefont{S.}~\bibnamefont{Neukirch}},
  \bibinfo{journal}{Journal of the Mechanics and Physics of Solids}
  \textbf{\bibinfo{volume}{69}}, \bibinfo{pages}{143} (\bibinfo{year}{2014}).

\bibitem[{\citenamefont{Oshri et~al.}(2015)\citenamefont{Oshri, Brau, and
  Diamant}}]{Oshri:2015bi}
\bibinfo{author}{\bibfnamefont{O.}~\bibnamefont{Oshri}},
  \bibinfo{author}{\bibfnamefont{F.}~\bibnamefont{Brau}}, \bibnamefont{and}
  \bibinfo{author}{\bibfnamefont{H.}~\bibnamefont{Diamant}},
  \bibinfo{journal}{Phys. Rev. E} \textbf{\bibinfo{volume}{91}},
  \bibinfo{pages}{052408} (\bibinfo{year}{2015}).

\bibitem[{\citenamefont{Hunt et~al.}(2000)\citenamefont{Hunt, Peletier,
  Champneys, Woods, Wadee, Budd, and Lord}}]{Hunt:D0NJhxAR}
\bibinfo{author}{\bibfnamefont{G.~W.} \bibnamefont{Hunt}},
  \bibinfo{author}{\bibfnamefont{M.~A.} \bibnamefont{Peletier}},
  \bibinfo{author}{\bibfnamefont{A.~R.} \bibnamefont{Champneys}},
  \bibinfo{author}{\bibfnamefont{P.~D.} \bibnamefont{Woods}},
  \bibinfo{author}{\bibfnamefont{M.~A.} \bibnamefont{Wadee}},
  \bibinfo{author}{\bibfnamefont{C.~J.} \bibnamefont{Budd}}, \bibnamefont{and}
  \bibinfo{author}{\bibfnamefont{G.~J.} \bibnamefont{Lord}},
  \bibinfo{journal}{Nonlinear Dynamics} \textbf{\bibinfo{volume}{21}},
  \bibinfo{pages}{3} (\bibinfo{year}{2000}).

\bibitem[{\citenamefont{Thompson and Champneys}(1996)}]{1996RSPSA.452..117T}
\bibinfo{author}{\bibfnamefont{J.~M.~T.} \bibnamefont{Thompson}}
  \bibnamefont{and} \bibinfo{author}{\bibfnamefont{A.~R.}
  \bibnamefont{Champneys}}, \bibinfo{journal}{Proc. R. Soc. London A}
  \textbf{\bibinfo{volume}{452}}, \bibinfo{pages}{117} (\bibinfo{year}{1996}).

\bibitem[{\citenamefont{Champneys and Thompson}(1996)}]{1996RSPSA.452.2467C}
\bibinfo{author}{\bibfnamefont{A.~R.} \bibnamefont{Champneys}}
  \bibnamefont{and} \bibinfo{author}{\bibfnamefont{J.~M.~T.}
  \bibnamefont{Thompson}}, \bibinfo{journal}{Proc. R. Soc. London A}
  \textbf{\bibinfo{volume}{452}}, \bibinfo{pages}{2467} (\bibinfo{year}{1996}).

\bibitem[{\citenamefont{Champneys and Groves}(1997)}]{1997JFM...342..199C}
\bibinfo{author}{\bibfnamefont{A.~R.} \bibnamefont{Champneys}}
  \bibnamefont{and} \bibinfo{author}{\bibfnamefont{M.~D.}
  \bibnamefont{Groves}}, \bibinfo{journal}{J. Fluid Mech.}
  \textbf{\bibinfo{volume}{342}}, \bibinfo{pages}{199} (\bibinfo{year}{1997}).

\bibitem[{\citenamefont{Burke and Knobloch}(2006)}]{BurkeK:2006}
\bibinfo{author}{\bibfnamefont{J.}~\bibnamefont{Burke}} \bibnamefont{and}
  \bibinfo{author}{\bibfnamefont{E.}~\bibnamefont{Knobloch}},
  \bibinfo{journal}{Phys. Rev. E} \textbf{\bibinfo{volume}{73}},
  \bibinfo{pages}{056211} (\bibinfo{year}{2006}).

\bibitem[{\citenamefont{Knobloch}(2015)}]{Knobloch:2015de}
\bibinfo{author}{\bibfnamefont{E.}~\bibnamefont{Knobloch}},
  \bibinfo{journal}{Annu. Rev. Condens. Matter Phys.}
  \textbf{\bibinfo{volume}{6}}, \bibinfo{pages}{325} (\bibinfo{year}{2015}).

\bibitem[{\citenamefont{Iooss and P{\'e}rou{\`e}me}(1993)}]{Iooss}
\bibinfo{author}{\bibfnamefont{G.}~\bibnamefont{Iooss}} \bibnamefont{and}
  \bibinfo{author}{\bibfnamefont{M.}~\bibnamefont{P{\'e}rou{\`e}me}},
  \bibinfo{journal}{J. Diff. Eq.} \textbf{\bibinfo{volume}{102}},
  \bibinfo{pages}{62} (\bibinfo{year}{1993}).

\bibitem[{\citenamefont{Kozyreff and Chapman}(2006)}]{KZ1}
\bibinfo{author}{\bibfnamefont{G.}~\bibnamefont{Kozyreff}} \bibnamefont{and}
  \bibinfo{author}{\bibfnamefont{S.~J.} \bibnamefont{Chapman}},
  \bibinfo{journal}{Phys. Rev. Lett.} \textbf{\bibinfo{volume}{97}},
  \bibinfo{pages}{044502} (\bibinfo{year}{2006}).

\bibitem[{\citenamefont{Chapman and Kozyreff}(2009)}]{KZ2}
\bibinfo{author}{\bibfnamefont{S.~J.} \bibnamefont{Chapman}} \bibnamefont{and}
  \bibinfo{author}{\bibfnamefont{G.}~\bibnamefont{Kozyreff}},
  \bibinfo{journal}{Physica D} \textbf{\bibinfo{volume}{238}},
  \bibinfo{pages}{319} (\bibinfo{year}{2009}).

\bibitem[{\citenamefont{Gaivão and Gelfreich}(2011)}]{Gelfreich2011}
\bibinfo{author}{\bibfnamefont{J.~P.} \bibnamefont{Gaivão}} \bibnamefont{and}
  \bibinfo{author}{\bibfnamefont{V.}~\bibnamefont{Gelfreich}},
  \bibinfo{journal}{Nonlinearity} \textbf{\bibinfo{volume}{24}},
  \bibinfo{pages}{677} (\bibinfo{year}{2011}).

\bibitem[{\citenamefont{Bergeon et~al.}(2008)\citenamefont{Bergeon, Burke,
  Knobloch, and Mercader}}]{Bergeon:2008fi}
\bibinfo{author}{\bibfnamefont{A.}~\bibnamefont{Bergeon}},
  \bibinfo{author}{\bibfnamefont{J.}~\bibnamefont{Burke}},
  \bibinfo{author}{\bibfnamefont{E.}~\bibnamefont{Knobloch}}, \bibnamefont{and}
  \bibinfo{author}{\bibfnamefont{I.}~\bibnamefont{Mercader}},
  \bibinfo{journal}{Phys. Rev. E} \textbf{\bibinfo{volume}{78}},
  \bibinfo{pages}{046201} (\bibinfo{year}{2008}).

\bibitem[{\citenamefont{Burke and Knobloch}(2009)}]{BurkeK:2009}
\bibinfo{author}{\bibfnamefont{J.}~\bibnamefont{Burke}} \bibnamefont{and}
  \bibinfo{author}{\bibfnamefont{E.}~\bibnamefont{Knobloch}},
  \bibinfo{journal}{Discrete and Continuous Dyn. Syst.-Suppl.}
  \textbf{\bibinfo{volume}{September}}, \bibinfo{pages}{109}
  (\bibinfo{year}{2009}).

\bibitem[{\citenamefont{Peletier and Troy}(2006)}]{Peletier}
\bibinfo{author}{\bibfnamefont{L.~A.} \bibnamefont{Peletier}} \bibnamefont{and}
  \bibinfo{author}{\bibfnamefont{W.~C.} \bibnamefont{Troy}},
  \emph{\bibinfo{title}{Spatial Patterns: Higher Order Models in Physics and
  Mechanics}} (\bibinfo{publisher}{Basel: Birkh\"{a}user},
  \bibinfo{year}{2006}).

\bibitem[{\citenamefont{Burke and Knobloch}(2007{\natexlab{a}})}]{BurkeK:2007a}
\bibinfo{author}{\bibfnamefont{J.}~\bibnamefont{Burke}} \bibnamefont{and}
  \bibinfo{author}{\bibfnamefont{E.}~\bibnamefont{Knobloch}},
  \bibinfo{journal}{Phys. Lett. A} \textbf{\bibinfo{volume}{360}},
  \bibinfo{pages}{681} (\bibinfo{year}{2007}{\natexlab{a}}).

\bibitem[{\citenamefont{Burke and Knobloch}(2007{\natexlab{b}})}]{BurkeK:2007b}
\bibinfo{author}{\bibfnamefont{J.}~\bibnamefont{Burke}} \bibnamefont{and}
  \bibinfo{author}{\bibfnamefont{E.}~\bibnamefont{Knobloch}},
  \bibinfo{journal}{Chaos} \textbf{\bibinfo{volume}{17}},
  \bibinfo{pages}{037102} (\bibinfo{year}{2007}{\natexlab{b}}).

\bibitem[{\citenamefont{Rivetti}(2013)}]{Rivetti:2013kk}
\bibinfo{author}{\bibfnamefont{M.}~\bibnamefont{Rivetti}},
  \bibinfo{journal}{Comptes Rendus M{\'e}canique}
  \textbf{\bibinfo{volume}{341}}, \bibinfo{pages}{333} (\bibinfo{year}{2013}).

\bibitem[{\citenamefont{Marple et~al.}(2015)\citenamefont{Marple, Purohit, and
  Veerapaneni}}]{Marple:2015kq}
\bibinfo{author}{\bibfnamefont{G.~R.} \bibnamefont{Marple}},
  \bibinfo{author}{\bibfnamefont{P.~K.} \bibnamefont{Purohit}},
  \bibnamefont{and}
  \bibinfo{author}{\bibfnamefont{S.}~\bibnamefont{Veerapaneni}},
  \bibinfo{journal}{Phys. Rev. E} \textbf{\bibinfo{volume}{92}},
  \bibinfo{pages}{012405} (\bibinfo{year}{2015}).

\bibitem[{\citenamefont{Cerda and Mahadevan}(2003)}]{Cerda:2003go}
\bibinfo{author}{\bibfnamefont{E.}~\bibnamefont{Cerda}} \bibnamefont{and}
  \bibinfo{author}{\bibfnamefont{L.}~\bibnamefont{Mahadevan}},
  \bibinfo{journal}{Phys. Rev. Lett.} \textbf{\bibinfo{volume}{90}},
  \bibinfo{pages}{074302} (\bibinfo{year}{2003}).

\bibitem[{\citenamefont{Brau et~al.}(2013)\citenamefont{Brau, Damman, Diamant,
  and Witten}}]{Brau:2013jn}
\bibinfo{author}{\bibfnamefont{F.}~\bibnamefont{Brau}},
  \bibinfo{author}{\bibfnamefont{P.}~\bibnamefont{Damman}},
  \bibinfo{author}{\bibfnamefont{H.}~\bibnamefont{Diamant}}, \bibnamefont{and}
  \bibinfo{author}{\bibfnamefont{T.~A.} \bibnamefont{Witten}},
  \bibinfo{journal}{Soft Matter} \textbf{\bibinfo{volume}{9}},
  \bibinfo{pages}{8177} (\bibinfo{year}{2013}).

\bibitem[{\citenamefont{Doedel et~al.}(1991{\natexlab{a}})\citenamefont{Doedel,
  Keller, and Kernevez}}]{1991IJBC....1..493D}
\bibinfo{author}{\bibfnamefont{E.}~\bibnamefont{Doedel}},
  \bibinfo{author}{\bibfnamefont{H.~B.} \bibnamefont{Keller}},
  \bibnamefont{and} \bibinfo{author}{\bibfnamefont{J.~P.}
  \bibnamefont{Kernevez}}, \bibinfo{journal}{Int. J. Bifurcat. Chaos}
  \textbf{\bibinfo{volume}{1}}, \bibinfo{pages}{493}
  (\bibinfo{year}{1991}{\natexlab{a}}).

\bibitem[{\citenamefont{Doedel et~al.}(1991{\natexlab{b}})\citenamefont{Doedel,
  Keller, and Kernevez}}]{1991IJBC....1..745D}
\bibinfo{author}{\bibfnamefont{E.}~\bibnamefont{Doedel}},
  \bibinfo{author}{\bibfnamefont{H.~B.} \bibnamefont{Keller}},
  \bibnamefont{and} \bibinfo{author}{\bibfnamefont{J.~P.}
  \bibnamefont{Kernevez}}, \bibinfo{journal}{Int. J. Bifurcat. Chaos}
  \textbf{\bibinfo{volume}{1}}, \bibinfo{pages}{745}
  (\bibinfo{year}{1991}{\natexlab{b}}).

\bibitem[{\citenamefont{Doedel et~al.}(2008)\citenamefont{Doedel, Champneys,
  Dercole, Fairgrieve, Kuznetsov, Oldeman, Paffenroth, Sandstede, Wang, and
  Zhang}}]{doedel08auto-07p}
\bibinfo{author}{\bibfnamefont{E.~J.} \bibnamefont{Doedel}},
  \bibinfo{author}{\bibfnamefont{A.~R.} \bibnamefont{Champneys}},
  \bibinfo{author}{\bibfnamefont{F.}~\bibnamefont{Dercole}},
  \bibinfo{author}{\bibfnamefont{T.}~\bibnamefont{Fairgrieve}},
  \bibinfo{author}{\bibfnamefont{Y.}~\bibnamefont{Kuznetsov}},
  \bibinfo{author}{\bibfnamefont{B.}~\bibnamefont{Oldeman}},
  \bibinfo{author}{\bibfnamefont{R.}~\bibnamefont{Paffenroth}},
  \bibinfo{author}{\bibfnamefont{B.}~\bibnamefont{Sandstede}},
  \bibinfo{author}{\bibfnamefont{X.}~\bibnamefont{Wang}}, \bibnamefont{and}
  \bibinfo{author}{\bibfnamefont{C.}~\bibnamefont{Zhang}},
  \emph{\bibinfo{title}{{AUTO-07P: Continuation and Bifurcation Software for
  Ordinary Differential Equations}}} (\bibinfo{year}{2008}).

\bibitem[{\citenamefont{Brau et~al.}(2010)\citenamefont{Brau, Vandeparre,
  Sabbah, Poulard, Boudaoud, and Damman}}]{Brau:2010fia}
\bibinfo{author}{\bibfnamefont{F.}~\bibnamefont{Brau}},
  \bibinfo{author}{\bibfnamefont{H.}~\bibnamefont{Vandeparre}},
  \bibinfo{author}{\bibfnamefont{A.}~\bibnamefont{Sabbah}},
  \bibinfo{author}{\bibfnamefont{C.}~\bibnamefont{Poulard}},
  \bibinfo{author}{\bibfnamefont{A.}~\bibnamefont{Boudaoud}}, \bibnamefont{and}
  \bibinfo{author}{\bibfnamefont{P.}~\bibnamefont{Damman}},
  \bibinfo{journal}{Nat. Phys.} \textbf{\bibinfo{volume}{7}},
  \bibinfo{pages}{56} (\bibinfo{year}{2010}).

\bibitem[{\citenamefont{{A. C. Chiang} and {K.
  Wainwright}}(2013)}]{chiang2013fundamental}
\bibinfo{author}{\bibnamefont{{A. C. Chiang}}} \bibnamefont{and}
  \bibinfo{author}{\bibnamefont{{K. Wainwright}}},
  \emph{\bibinfo{title}{Fundamental Methods of Mathematical Economics}}
  (\bibinfo{publisher}{McGraw-Hill/Irwin}, \bibinfo{address}{Boston, Mass.},
  \bibinfo{year}{2013}), \bibinfo{edition}{4th} ed.

\bibitem[{\citenamefont{Diamant and Witten}(2013)}]{Diamant:2013je}
\bibinfo{author}{\bibfnamefont{H.}~\bibnamefont{Diamant}} \bibnamefont{and}
  \bibinfo{author}{\bibfnamefont{T.~A.} \bibnamefont{Witten}},
  \bibinfo{journal}{Phys. Rev. E} \textbf{\bibinfo{volume}{88}},
  \bibinfo{pages}{012401} (\bibinfo{year}{2013}).

\bibitem[{\citenamefont{Paulsen et~al.}(2016)\citenamefont{Paulsen, Hohlfeld,
  King, Huang, Qiu, Russell, Menon, Vella, and Davidovitch}}]{Paulsen:2016hm}
\bibinfo{author}{\bibfnamefont{J.~D.} \bibnamefont{Paulsen}},
  \bibinfo{author}{\bibfnamefont{E.}~\bibnamefont{Hohlfeld}},
  \bibinfo{author}{\bibfnamefont{H.}~\bibnamefont{King}},
  \bibinfo{author}{\bibfnamefont{J.}~\bibnamefont{Huang}},
  \bibinfo{author}{\bibfnamefont{Z.}~\bibnamefont{Qiu}},
  \bibinfo{author}{\bibfnamefont{T.~P.} \bibnamefont{Russell}},
  \bibinfo{author}{\bibfnamefont{N.}~\bibnamefont{Menon}},
  \bibinfo{author}{\bibfnamefont{D.}~\bibnamefont{Vella}}, \bibnamefont{and}
  \bibinfo{author}{\bibfnamefont{B.}~\bibnamefont{Davidovitch}},
  \bibinfo{journal}{Proc. Nat. Acad. Sci.} \textbf{\bibinfo{volume}{113}},
  \bibinfo{pages}{1144} (\bibinfo{year}{2016}).

\bibitem[{\citenamefont{Paulsen et~al.}(2017)\citenamefont{Paulsen, D{\'e}mery,
  Toga, Qiu, Russell, Davidovitch, and Menon}}]{Paulsen:2017dq}
\bibinfo{author}{\bibfnamefont{J.~D.} \bibnamefont{Paulsen}},
  \bibinfo{author}{\bibfnamefont{V.}~\bibnamefont{D{\'e}mery}},
  \bibinfo{author}{\bibfnamefont{K.~B.} \bibnamefont{Toga}},
  \bibinfo{author}{\bibfnamefont{Z.}~\bibnamefont{Qiu}},
  \bibinfo{author}{\bibfnamefont{T.~P.} \bibnamefont{Russell}},
  \bibinfo{author}{\bibfnamefont{B.}~\bibnamefont{Davidovitch}},
  \bibnamefont{and} \bibinfo{author}{\bibfnamefont{N.}~\bibnamefont{Menon}},
  \bibinfo{journal}{Phys. Rev. Lett.} \textbf{\bibinfo{volume}{118}},
  \bibinfo{pages}{048004} (\bibinfo{year}{2017}).

\bibitem[{\citenamefont{Gopal et~al.}(2006)\citenamefont{Gopal, Belyi, Diamant,
  Witten, and Lee}}]{Gopal:2006gk}
\bibinfo{author}{\bibfnamefont{A.}~\bibnamefont{Gopal}},
  \bibinfo{author}{\bibfnamefont{V.~A.} \bibnamefont{Belyi}},
  \bibinfo{author}{\bibfnamefont{H.}~\bibnamefont{Diamant}},
  \bibinfo{author}{\bibfnamefont{T.~A.} \bibnamefont{Witten}},
  \bibnamefont{and} \bibinfo{author}{\bibfnamefont{K.~Y.~C.}
  \bibnamefont{Lee}}, \bibinfo{journal}{J. Phys. Chem. B}
  \textbf{\bibinfo{volume}{110}}, \bibinfo{pages}{10220}
  (\bibinfo{year}{2006}).

\bibitem[{\citenamefont{Leahy et~al.}(2010)\citenamefont{Leahy, Pocivavsek,
  Meron, Lam, Salas, Viccaro, Lee, and Lin}}]{Leahy:2010he}
\bibinfo{author}{\bibfnamefont{B.~D.} \bibnamefont{Leahy}},
  \bibinfo{author}{\bibfnamefont{L.}~\bibnamefont{Pocivavsek}},
  \bibinfo{author}{\bibfnamefont{M.}~\bibnamefont{Meron}},
  \bibinfo{author}{\bibfnamefont{K.~L.} \bibnamefont{Lam}},
  \bibinfo{author}{\bibfnamefont{D.}~\bibnamefont{Salas}},
  \bibinfo{author}{\bibfnamefont{P.~J.} \bibnamefont{Viccaro}},
  \bibinfo{author}{\bibfnamefont{K.~Y.~C.} \bibnamefont{Lee}},
  \bibnamefont{and} \bibinfo{author}{\bibfnamefont{B.}~\bibnamefont{Lin}},
  \bibinfo{journal}{Phys. Rev. Lett.} \textbf{\bibinfo{volume}{105}},
  \bibinfo{pages}{058301} (\bibinfo{year}{2010}).

\bibitem[{\citenamefont{Jambon-Puillet
  et~al.}(2017)\citenamefont{Jambon-Puillet, Josserand, and
  Proti{\`e}re}}]{JambonPuillet:2017fa}
\bibinfo{author}{\bibfnamefont{E.}~\bibnamefont{Jambon-Puillet}},
  \bibinfo{author}{\bibfnamefont{C.}~\bibnamefont{Josserand}},
  \bibnamefont{and}
  \bibinfo{author}{\bibfnamefont{S.}~\bibnamefont{Proti{\`e}re}},
  \bibinfo{journal}{Phys. Rev. Materials} \textbf{\bibinfo{volume}{1}},
  \bibinfo{pages}{042601} (\bibinfo{year}{2017}).

\end{thebibliography}

\end{document}